\documentclass[%
 reprint,
superscriptaddress,
nofootinbib,
 amsmath,amssymb,
 aps,
pra,
]{revtex4-2}

\usepackage{graphicx}
\usepackage{xcolor}
\usepackage[normalem]{ulem}
\usepackage{dcolumn}
\usepackage{bm}

\usepackage[colorlinks=true,
            linkcolor=magenta,
            citecolor=cyan,
            filecolor=magenta,
            urlcolor=magenta]{hyperref}
\usepackage{aas_macros}
\usepackage{mathrsfs}
\usepackage{multirow}
\usepackage{pifont}
\newcommand{\cmark}{\ding{51}}%
\newcommand{\xmark}{\ding{55}}%

\begin{document}

\preprint{APS/123-QED}

\title{Expectations for the first supermassive black-hole binary resolved by PTAs\\ I: Model efficacy}

\author{Levi Schult}
\email[Corresponding author: ]{levi.s.schult@vanderbilt.edu}
\affiliation{Department of Physics and Astronomy, Vanderbilt University, 2301 Vanderbilt Place, Nashville, TN 37235, USA}

\author{Polina Petrov}%
\affiliation{Department of Physics and Astronomy, Vanderbilt University, 2301 Vanderbilt Place, Nashville, TN 37235, USA}

\author{Stephen~R.~Taylor}
\affiliation{Department of Physics and Astronomy, Vanderbilt University, 2301 Vanderbilt Place, Nashville, TN 37235, USA}

\author{Nihan Pol}
\affiliation{Department of Physics \& Astronomy, Texas Tech University, Lubbock, TX, 79409, USA}

\author{Nima Laal}
\affiliation{Department of Physics and Astronomy, Vanderbilt University, 2301 Vanderbilt Place, Nashville, TN 37235, USA}

\author{Maria Charisi}
\affiliation{Department of Physics and Astronomy, Washington State University, Pullman, WA 99163, USA}
\affiliation{Institute of Astrophysics, FORTH, GR-71110, Heraklion, Greece}

\author{Chung-Pei Ma}
\affiliation{Department of Astronomy, University of California, Berkeley, 501 Campbell Hall 3411, Berkeley, CA 94720, USA}
\affiliation{Department of Physics, University of California, Berkeley, CA 94720, USA}

\begin{abstract}
One of the most promising targets for Pulsar Timing Arrays (PTAs) is identifying an individual supermassive black hole binary (SMBHB) out of the population of binaries theorized to produce a gravitational wave background (GWB). In this work, we emulate an evolving PTA dataset, complete with an increasing number of pulsars and timing baseline, into which we inject a single binary on top of a Gaussian GWB signal. We vary the binary's source parameters, including sky position and frequency, and create ten noise realizations for each source/PTA combination to synthesize an ensemble of simulated datasets with which we assess current Bayesian binary search techniques. We develop a novel, cross-correlation based model tuned for individual SMBHBs and compare its binary detection and characterization capabilities to two waveform-based template models. This Spike Pixel (SP) technique is designed to identify the frequency-specific anisotropy induced by an individual SMBHB.
We find that a template-based search including the full gravitational-wave signal structure (i.e., both the Earth and pulsar effects of an incident GW) returns the highest Bayes Factors (BF) and has the most robust parameter estimation. SP attains a realization-median BF $> 10$ at source strengths $7<$(S/N)$<15$. Interestingly, despite being a deterministic model, the Earth-term template struggles to identify and characterize low-frequency binaries (i.e., at $5 ~ \mathrm{nHz}$). These binaries require higher source strengths ($16<$ (S/N) $<19$) to reach the same BF threshold. This is likely due to neglected confusion effects between the pulsar and Earth terms. By contrast, SP shows promise for parameter estimation despite treating a binary's GW signal as excess directional GW power without phase modeling. Sky location and frequency parameter constraints returned by SP are only surpassed by the Earth term template model at (S/N) $\sim 12-13$. Milestones for a first detection using the full-signal GW model are included in a companion paper \citet{2025arXiv251001316P}.
\end{abstract}

\maketitle


\section{\label{sec:Intro}Introduction}

The extraordinary timekeeping abilities of millisecond pulsars enable the study of low-frequency gravitational waves (GWs) through Pulsar Timing Arrays (PTAs) \cite{1979ApJ...234.1100D, 1978SvA....22...36S}. By building sophisticated deterministic models to predict the arrival time of electromagnetic (EM) radiation from the pulsar at a radio telescope, one can gain physical intuition about processes that induce unmodeled delays. These delays can be the result of astrometric uncertainties, binary companions, the ionized interstellar medium, and, importantly, GWs \citep{condon2016essential, lorimer2005handbook}. The primary expected source of low-frequency GWs targeted by PTAs are inspiraling supermassive black hole binaries (SMBHBs). Due to the long evolution times of these systems (thousands to millions of years), an individual binary is a monochromatic, continuous GW (CW) that will persist in PTA datasets for the foreseeable future \cite{2010arXiv1008.1782C}.

These binaries form when galaxies, hosting massive central SMBHs, merge over cosmic time \cite{1980Natur.287..307B}. Dynamical friction brings the BHs to the center of the newly formed galaxy before stellar three-body and gas interactions extract angular momentum from the system, bringing the binary to a sub-parsec scale. At these small orbital separations, GWs are the primary mechanism driving the BHs to merge.  The timescales and efficiency of each process are currently open areas of research \citep{2025arXiv250814253H}. Early studies suggested that stellar three-body interactions would not be sufficient to harden the binary orbit to tight separations, giving rise to a ``final parsec problem" of stalled binaries \cite{2003ApJ...596..860M, 2019A&ARv..27....5B} (See Section 3.2.2 of \citet{2024arXiv240207571C}). However, many possible solutions have been proposed, including the presence of nuclear star clusters \citep{2025ApJ...981..203M}, stellar bulge triaxiality and galaxy structure \citep{2022MNRAS.512.3365B, 2013ApJ...773..100K}, and galactic rotation \citep{2017MNRAS.470..940M}. 

Moreover, if the binary is embedded in a disk with torquing that helps to shrink the orbit, then a byproduct may be electromagnetic (EM) brightness variations linked to the binary's orbital period and (potentially) phase. This presents an SMBHB as a near-indefinite multimessenger system due to the long timescale of SMBHB inspiral \citep{2022LRR....25....3B, 2023arXiv231016896D}. With both the GW signal from an individual SMBHB and EM observations of the host galaxy and system, a full picture of the binary's evolution can be constructed. Further, multimessenger observations would also reveal answers to other existing open questions about accretion around SMBHBs and how central BHs coevolve with their host galaxies \cite{2019BAAS...51c.490K}. Observations of sub-parsec-separated and merging SMBHBs are needed to adjudicate the most influential of these mechanisms.

The population of millions of these SMBHBs formed by the cosmic history of galaxy mergers, all emitting GWs that overlap incoherently, produces a stochastic GW background (GWB). This GWB subsumes and can obscure individual binaries from detection. It has been the main target for PTA experiments for decades, as it has long been theorized to be the first signal to be detected \citep{2015MNRAS.451.2417R, 2024ApJ...965..164G}. Its telltale signature, distinguishing it from other noise sources present in PTAs such as solar wind or clock errors \citep{2016MNRAS.455.4339T}, is a quadrupolar-like spatial correlation across pulsars in the array, known as the Hellings-Downs (HD) curve \cite{1983ApJ...265L..39H, 2016MNRAS.455.4339T}. PTA experiments around the world have recently found evidence for these correlations to varying degrees of significance \cite{2024ApJ...966..105A, 2023ApJ...951L...8A, 2024A&A...685A..94E, 2023ApJ...951L...6R, 2023RAA....23g5024X, 2025MNRAS.536.1489M}. The HD curve is not uniquely indicative of SMBHBs as the source, but rather a fingerprint of a statistically isotropic GWB. While the GWB spectrum is consistent with predictions of populations of SMBHBs \cite{2023ApJ...952L..37A, 2024A&A...685A..94E}, there are other theorized origins of the GWB \cite{2023ApJ...951L..11A, 2024A&A...685A..94E}. Spectral characterization can provide population-level information about environmental influences on SMBHB evolution, but these are indirect inferences. The direct multimessenger observation of an individual SMBHB and its host galaxy can help cut the Gordian knot of questions surrounding the confluence of mechanisms involved in SMBHB dynamics and demographics in galaxy centers. 

In this work, we focus on resolving the GW signal of an individual SMBHB out of the confusion GWB posed by the rest of the binary population. We employ a variety of methods, comparing their efficacy, with attention focused on the key system parameters that are required to facilitate host galaxy identification. These different GW Bayesian analysis techniques fall into three main categories: Fourier-domain \citep{2015MNRAS.449.1650Z, 2024arXiv241213379G}, time-domain template-based \citep{2022PhRvD.105l2003B, 2013CQGra..30v4004E, 2024CQGra..41v5017B}, and power-based searches \citep{2020PhRvD.102h4039T, 2018MNRAS.477.5447G}. 
Fourier-domain approaches identify the imprint of an individual binary via its effect on the Fourier coefficients that describe the pulsar timing residuals. Alternatively, power-based techniques search instead for excess GW power on the sky, creating deviations from an isotropic GWB and therefore the induced timing residuals.

Template-based approaches use deterministic GW waveforms from parameters describing the binary such as the chirp mass, sky location, and inclination angle. These searches can be extremely complex due to the fact that effects of the GW are seen at the Earth and the pulsar. This creates a pulsar-coherent signal (Earth term) and pulsar-specific signal (pulsar term), requiring model parameters for each pulsar in the array (further explanation can be found in \autoref{sec:methods_gwsfromabhb}) \cite{2012PhRvD..86l4028B, 2010arXiv1008.1782C, 2013CQGra..30v4004E}. One can also ignore the pulsar terms, vastly shrinking the parameter space at the expense of model accuracy.\footnote{\citet{2024PhRvL.132f1401C} have found that targeted (fixed sky location and distance signal models) Earth-term searches do not suffer greatly diminished performance.} While this reduces computational demand, model mis-specification can abound, resulting in biased parameter estimations \cite{Zhu2016,2024arXiv240721105F}. While detection of a discrete binary using a deterministic model would be definitive, it remains to be seen whether these approaches are capable of resolving the binary before a power-based search. This outstanding question from \citet{2020PhRvD.102h4039T} motivated this work and inspired us to develop a specialized anisotropy model to search for a binary via its cross correlations. 
We present this novel technique in \autoref{sec:Methods}.

The estimation of the binary's frequency and sky location are critical parameters for host galaxy identification and EM followup (See \cite{2023arXiv231016896D} for a review of SMBHB EM counterparts).\footnote{Measurements of the binary chirp mass are also important for host galaxy identification \citep{2024ApJ...976..129P}, but only a full signal waveform model returns such constraints. The other two models tested in this work only return combined strain or power parameters containing both the chirp mass and luminosity distance.} Since PTA localizations are expected to be large \citep{2024ApJ...976..129P, 2025arXiv250401074T, SV2010, Taylor2016, G19}, knowledge of which model robustly recovers these parameters will aid in accelerating these campaigns. One can imagine that these parameters could be estimated earlier using simpler models rather than a deterministic technique. One such model is a frequency-resolved search for anisotropy in the GWB. These power-based searches highlight regions of excess GW power on the sky caused by especially loud binaries, rising above the GWB.\footnote{While one might expect that GW anisotropy is due to the distribution of galaxies on the sky, the prevailing reason is due to the Poisson statistics governing the distance to members of the SMBHB population \cite{2024arXiv240616031A, 2024arXiv240405670G, 2024arXiv241108744G}.} The anisotropy induced by individual SMBHBs creates a tantalizing prospect for identifying a potential binary via a hotspot of GW power on the sky \cite{2020PhRvD.102h4039T, 2024ApJ...965..164G}. 

This work aims to answer whether such a power-based search could identify an individual binary before other approaches. Further, how well can each method characterize the binary's key parameters necessary for host galaxy identification (sky location and frequency)? The evolution of the binary's parameter estimation through successive PTA datasets is explored in \citet{2025arXiv251001316P} (hereafter Paper II).

This paper is organized as follows. We present our various SMBHB search models in \autoref{sec:Methods}, and the details of the simulated datasets to which they are applied in \autoref{sec:simdatasets}. We then summarize the ensemble efficacy of these models in the various simulations in \autoref{sec:Results}. Further discussion of these results is provided in \autoref{sec:Disc}, along with concluding remarks in \autoref{sec:conclusion}. Throughout this work we use natural units where $G = c = 1$.

\section{\label{sec:Methods} Methods}
\subsection{GWs from a Binary}
\label{sec:methods_gwsfromabhb}
The GW-induced shift to the arrival rate of radio pulses from pulsars can be written as
\begin{equation}
\label{eqn:pulseredshift}
    z(t, \hat{\Omega}) = \frac{1}{2}\frac{\hat{p}^k\hat{p}^l}{1+\hat{\Omega}\cdot\hat{p}}\Delta h_{kl}.
\end{equation}
where $\hat{\Omega}$ is the direction of GW propagation, and $\hat{p}$ is the unit vector pointing from the Earth (or more accurately, the solar system barycenter) to the pulsar.
The term $\Delta h_{kl}$ encodes the difference in the metric perturbation at the time of passing the pulsar ($t_p$) and Earth ($t_e$), which is a consequence of the integrated effect of the radio pulse propagating through the perturbed spacetime:
\begin{equation} \label{eq:delta-h}
    \Delta h_{kl} = h_{kl}(t_p, \hat{\Omega}) -h_{kl}(t_e, \hat{\Omega}),
\end{equation}
where 
\begin{equation}
\label{eqn:pulsartime}
    t_p = t_e - L_{p}(1+\hat{\Omega}\cdot\hat{p})
\end{equation}
and $L_p$ is the distance to the pulsar. 

The perturbations themselves, $h_{kl}$, can be written as a sum over the plus and cross GW polarizations,
\begin{equation}
\label{eqn:metricperturbation}
    h_{kl}(t, \hat{\Omega}) = e_{kl}^{+}(\hat{\Omega})h_+(t, \hat{\Omega}) + e^{\times}_{kl}(\hat{\Omega})h_\times(t, \hat{\Omega}).
\end{equation}
where $h_{+,\times}$ and $e_{kl}^{+, \times}$ are the polarization amplitude and basis tensors, respectively. These basis tensors can be described in terms of outer products of orthonormal vectors in the plane perpendicular to $\hat\Omega$ \cite{1987GReGr..19.1101W},
\begin{align}
    e^+_{kl} &= \hat{m}_k\hat{m}_l - \hat{n}_k\hat{n}_l, \nonumber\\
    e^{\times}_{kl} &= \hat{m}_k\hat{n}_l + \hat{n}_k\hat{m}_l,
\end{align}
which form a right-handed basis triad:
\begin{align}
    \hat{\Omega} &= -(\sin\theta\cos\phi)\hat{x} - (\sin\theta\sin\phi)\hat{y} - (\cos\theta)\hat{z} \nonumber \\
    \hat{m} &= (\sin\phi)\hat{x} - (\cos\phi)\hat{y}\\
    \hat{n} &= -(\cos\theta\cos\phi)\hat{x} - (\cos\theta\sin\phi)\hat{y} + (\sin\theta)\hat{z},
\end{align}
where $\theta = \pi/2 - \delta$ and $\phi = \alpha$ are the polar and azimuthal coordinate angles of the GW source ($\alpha, \delta$ are the equatorial right ascension and declination coordinates of the source). 

Our observables are not shifts in the arrival rate of pulses, but rather times of arrival (TOAs) themselves, from which we can infer unmodeled delays or offsets, $s(t)$. The latter are modeled as a time integral over $z(t)$, such that
\begin{align}
    s(t) &= \int_0^t z(t')dt' \nonumber\\
         &= \int_0^t F^A(\hat{\Omega})\Delta h_A(t', \hat{\Omega}) dt' \nonumber\\
         &= F^+(\hat{\Omega})\Delta s_+(t) + F^\times (\hat{\Omega}) \Delta s_\times(t).
\end{align}
where
\begin{equation}
        F^A(\hat{\Omega}) = \frac{1}{2} \frac{\hat{p}^k \hat{p}^l}{1+\hat{\Omega}\cdot\hat{p}}e^A_{kl},
\end{equation}
is the \textit{GW antenna response function} of a particular Earth-pulsar system, and
\begin{equation}
    \Delta s_{A}(t) = \int_0^t \Delta h_A(t', \hat{\Omega}) dt'.
\end{equation}

Our goal is to model the GW signal of a single binary on top of a GWB from the remainder of the SMBHB population. We can model this situation either with a deterministic signal model or through the framework of GWB anisotropy. In the former case, one models the CW individually by subtracting a deterministic GW signal to leave residuals whose covariance depends on the statistical properties of the GWB and noise. Alternatively, we can attempt to model both GW sources as a highly anisotropic GWB, where the single binary is a single-direction, single-frequency excess in the otherwise statistically isotropic GWB.

\subsubsection{Deterministic Model}

For a binary in a circular orbit with non-spinning BHs, the zeroth post-Newtonian order effect on TOAs has the following polarization-specific time offsets \cite{2012ApJ...756..175E, 2013CQGra..30v4004E}:
\begin{align}
\label{eqn:circcwresid}
    s_+(t) &= \frac{\mathcal{M}^{5/3}}{d_L \omega(t)^{1/3}} \left\{\sin[2(\Phi(t)](1+\cos^2\iota)\cos2\psi \right. \nonumber \\
    &\left.+ 2\cos[2\Phi(t)]\cos \iota \sin(2\psi) \right\} \\
    s_{\times}(t) &= \frac{\mathcal{M}^{5/3}}{d_L \omega(t)^{1/3}} \left\{-\sin[2\Phi(t)](1+\cos^2\iota)\sin(2\psi) \right. \nonumber\\
    &\left. + 2\cos[2\Phi(t)]\cos \iota \cos2\psi \right\}
\end{align}
where $\mathcal{M} = (m_1 m_2)^{3/5} / (m_1 + m_2)^{1/5}$ is the chirp mass, $\iota$ is the inclination angle between the observer's line-of-sight and the binary's orbital angular momentum vector, $d_L$ is the binary's luminosity distance, $\omega(t)$ is the orbital angular frequency, and $\Phi(t)$ is the orbital phase of the binary.\footnote{While the observed chirp mass and orbital frequency are redshifted from their rest-frame values ($\mathcal{M}_r = \mathcal{M}/(1+z)$ and $\omega_r = \omega_0(1+z)$, with $z$ being the redshift of the source), we focus only on local sources ($z\approx 0$).} Lastly, $\psi$ is the GW polarization angle. We do not search over the luminosity distance as a parameter of our model, instead opting to search over the strain amplitude of the binary, $h_0$. Using the GW frequency, related to our orbital angular frequency by $\omega(t)=\pi f_\mathrm{GW}(t)$, our strain is defined as \citep{2009LRR....12....2S},
\begin{equation}
    h_0 = \frac{2 \mathcal{M}^{5/3}(\pi f_\mathrm{GW}(t))^{2/3}}{d_L}.
\end{equation}

The GWs radiated by the binary evolve its angular phase and frequency, such that \citep{1964PhRv..136.1224P}
\begin{align}
    \Phi(t) &= \Phi_0 + \frac{1}{32}\mathcal{M}^{-5/3}[\omega_0^{-5/3} - \omega(t)^{-5/3}], \\
    \omega(t) &= \omega_0\left(1 - \frac{256}{5} \mathcal{M}^{5/3}\omega_0^{8/3}t\right)^{-3/8},
\end{align}
where $\Phi_0$ and $\omega_0$ are the initial angular phase and frequency at some reference time, $t_0$. We assume that the binary will not undergo appreciable frequency evolution over the 22 years of our simulated datasets, such that $\omega(t)\approx \omega_0$ and $\Phi(t)\approx \Phi_0 + \omega_0 (t - t_0)$. 

The pulsar term corresponds to an earlier stage in the binary's inspiral (see \autoref{eqn:pulsartime}), so the orbital frequency and phase must be evolved backward to leading order in our model:
\begin{align}
\label{eqn:pulsarphase}
    \Phi_p (t) &= \Phi_0 + \omega_0 t - \omega_0 L (1 + \hat{\Omega} \cdot \hat{p}) - \dot{\omega} L (1 + \hat{\Omega} \cdot \hat{p})t, \nonumber\\
    \omega_p (t) &= \omega_0 - \dot{\omega} L (1 + \hat{\Omega}\cdot\hat{p})
\end{align}

We follow the procedure from \citet{2013CQGra..30v4004E} to search over the pulsar term phase and pulsar distance as separate variables, even though we can see above that the phase is a derived quantity from the distance and other binary parameters. However, this makes the Markov-Chain Monte Carlo (MCMC) exploration of the parameter space far more efficient, since the PTA likelihood is otherwise highly oscillatory in the pulsar distance parameters. 

While this describes the full CW signal model, one can reduce the model complexity by neglecting the pulsar terms and only modeling the Earth-term signal. This can be a reasonable approximation when the pulsar terms sufficiently differ in frequency from the Earth term (e.g., by a typical frequency resolution bin) \citep{2012ApJ...756..175E}. This renders the signal---under the assumptions of negligible evolution over our PTA dataset duration---entirely monochromatic.

It also grants greater speed and efficiency to the MCMC exploration of the (now lower-dimensional) parameter space at the cost of model accuracy. However, recent work by \citet{2024arXiv240721105F} and \citet{2024PhRvL.132f1401C} show how, despite the inherent biases in an Earth-term-only search due to ignoring signal structure, they can return useful constraints on binary parameters. 

\subsubsection{Anisotropic GWB model for a single binary}

A single binary producing the entire GW signal in the PTA band is the most extreme scenario for GW anisotropy. In our case it is moderated by the presence of a Gaussian GWB. We begin by summarizing the framework for modeling a GWB in PTA data \citep[see, e.g.,][for more details]{2018gwv..book.....M,2021arXiv210513270T}, then elucidate how our single-direction, single-pixel model of binary-excess GWB power is constructed.

A general metric perturbation at $\vec{x}$ can be decomposed as a superposition over frequencies and propagation directions of plane GWs: 
\begin{equation} \label{eq:general-h}
    h_{kl}(t, \vec{x}) =\!\!\! \sum_{A=+, \times}\! \int_{-\infty}^\infty\!\! df\!\!  \int_{S^2} \!\!d \hat{\Omega}\, h_A(f, \hat{\Omega}) \, e^{-2 \pi i f (t-\hat{\Omega}\cdot\vec{x})} e^A_{kl}(\hat{\Omega}).
\end{equation}

With this notation, the $\Delta h_{kl}$ quantity in \autoref{eq:delta-h} can be formed by phase-lagging the complex exponential quantity in \autoref{eq:general-h} with $t_p = t - L(1+\hat{\Omega}\cdot\hat{p})$, such that
\begin{align}
\label{eqn:redshiftplanewave}
    z(t,\hat{\Omega}) = \!\!\!\! \sum_{A=+, \times} \!\int_{-\infty}^\infty &df  \!\int_{S^2} \!\! d\hat{\Omega}\, F^A(\hat{\Omega}) \, h_{A}(f, \hat{\Omega}) \nonumber\\
    &\times e^{-2 \pi i f t} [e^{2 \pi i f L_p(1+\hat{\Omega} \cdot \hat{p})} - 1].
\end{align}

A GWB is searched through the cross-correlations in pulsar timing residuals, related to the cross-correlated shifts to pulse arrival rates. We assume a stationary, Gaussian, and statistically unpolarized stochastic GWB signal, i.e.,
\begin{equation}
    \langle h^*_A(f, \hat{\Omega})h_{A'}(f', \hat{\Omega}')\rangle \!\!=\!\! \frac{\delta_{AA'}}{2}  \frac{\delta^2(\hat{\Omega},\hat{\Omega}')}{4\pi}\frac{\delta(f\!\!-\!\!f')}{2} S_h(f)P(\hat{\Omega}),
\end{equation}
where $S_h(f)$ is the one-sided Power Spectral Density (PSD) of the GWB and $P(\hat{\Omega})$ is the angular distribution of power on the sky. This  simplifies the expectation over ensembles for pulse redshift correlations between pulsars $a$ and $b$, such that
\begin{align}
\label{eqn:redshiftexpectation}
    \langle z^*_a(t)z_b(t)\rangle =  \frac{1}{2} \int_{-\infty}^{\infty} &df \, S_h(f) \int_{S^2} \frac{d^2 \hat{\Omega}}{8\pi}  \, \kappa_{ab}(f, \hat{\Omega}) \nonumber\\
    &\times P(\hat{\Omega}) \sum_{A=+,\times}\!\!F^A_a(\hat{\Omega})F^{A}_b(\hat{\Omega}),
\end{align}
where
\begin{equation}
    \kappa_{ab}(f, \hat{\Omega})\!  =\!
    \left[ e^{-2 \pi i f L_a (1+\hat{\Omega}\cdot\hat{p}_a)} - 1 \right] \!\left[ e^{2 \pi i f L_b (1+\hat{\Omega}\cdot\hat{p}_b)}  - 1 \right],
\end{equation}
encompasses the Earth and pulsar term wave interference effects. PTAs operate in the short-wavelength regime \cite{2024CQGra..41q5008R} with $fL\gtrsim 10$, such that $\kappa_{ab}$ is highly oscillatory and averages to $\kappa_{ab}\approx (1+\delta_{ab})$. The overlap reduction function (ORF) thus approximates the inclusion of pulsar term effects via a Kronecker delta function on pulsar indices. For a statistically isotropic GW sky, the ORF is given by assuming $P(\hat\Omega)=1\,\,\forall\,\,\hat\Omega$, such that
\begin{align}
\label{eqn:ORFformulation}
\tilde{\Gamma}_{ab} &= \int_{S^2} \frac{d^2\hat{\Omega}}{4 \pi} \kappa_{ab}\!\! \sum_{A=+, \times} \!\!F^A_a(\hat{\Omega})F^A_b(\hat{\Omega})
\\
    \Gamma^{\rm HD}_{ab} &= \frac{3}{2} \tilde{\Gamma}_{ab} \\
    \label{eqn:HDeqn}
    \Gamma^{\rm HD}_{ab} &= \frac{3}{2} x_{ab} \ln(x_{ab}) - \frac{1}{4}x_{ab} + \frac{1}{2} + \frac{1}{2}\delta_{ab}.
\end{align}
where $x_{ab} = (1-\cos\xi_{ab})/2$ and $\xi_{ab}$ is the angle between pulsars $a$ and $b$. We have included the customary prefactor in \autoref{eqn:HDeqn} to normalize the Hellings-Downs (HD) ORF to 1 for $a=b$.\footnote{While we have skipped the evaluation of the integral that results in the HD ORF, the interested reader can examine \citet{2009PhRvD..79h4030A, 2013PhRvD..88f2005M, 2014PhRvD..90h2001G} and \citet{2015AmJPh..83..635J}.}

We now write the one-sided PSD of the GWB in terms of the characteristic strain, $S_h(f) = h_c^2 / f$. We also account for the translation of pulse arrival rate into timing residual space, which gives an extra factor in the cross-correlations of $1/4 \pi^2f^2$ due to the time integrals. The cross-PSD of timing residuals induced by an isotropic GWB is then
\begin{equation}
    S_{ab}(f) = \frac{1}{3}\frac{S_h(f)}{4 \pi^2 f^2}\Gamma^{\rm HD}_{ab} = \frac{h_c^2(f)}{12 \pi^2 f^3}\Gamma^{\rm HD}_{ab}.
\end{equation}

Our goal with this approach is to model the emergence of a single SMBHB's GW signal from the confusion GWB of the rest of the binary population. We do so through GWB anisotropy; this anisotropy will be narrowband in frequency, associated with the binary's GW frequency, and as a point source on the sky. We thus introduce the \textit{Spike Pixel} (SP) model, for which the ORF corresponds to that of a single pixel in which the GW origin lies (similar to a radiometer search \cite{2006CQGra..23S.179B, 2008PhRvD..77d2002M, 2009PhRvD..80l2002T, 2020PhRvD.102h4039T, 2023ApJ...956L...3A}), with a PSD that is a single spike at the putative binary's Earth-term GW frequency. Using the normalization of the GW power, for a single point, we have
\begin{align}
   \int_{S^2} P(\hat{\Omega})d^2 \hat{\Omega}
    =\int_{S^2} P_{\hat{\Omega}'} \delta^2(\hat{\Omega}, \hat{\Omega}') d^2 \hat{\Omega} = 4\pi \\
    P_{\hat{\Omega}'} = 4\pi,
\end{align}
where $\hat\Omega'$ is the propagation direction of the putative GW signal, which is the negative of the unit vector pointing to the binary origin. The ORF of a single point of power is then
\begin{equation}
\label{eqn:pixorf}
    \aleph_{ab}(\hat\Omega') = \frac{3}{2} (1+\delta_{ab}) \!\!\sum_{A=+,\times} \!\! F^A_a(\hat{\Omega}')F^A_b(\hat{\Omega}').
\end{equation}

\begin{figure}[!ht]
    \centering
    \includegraphics[width=0.5\textwidth]{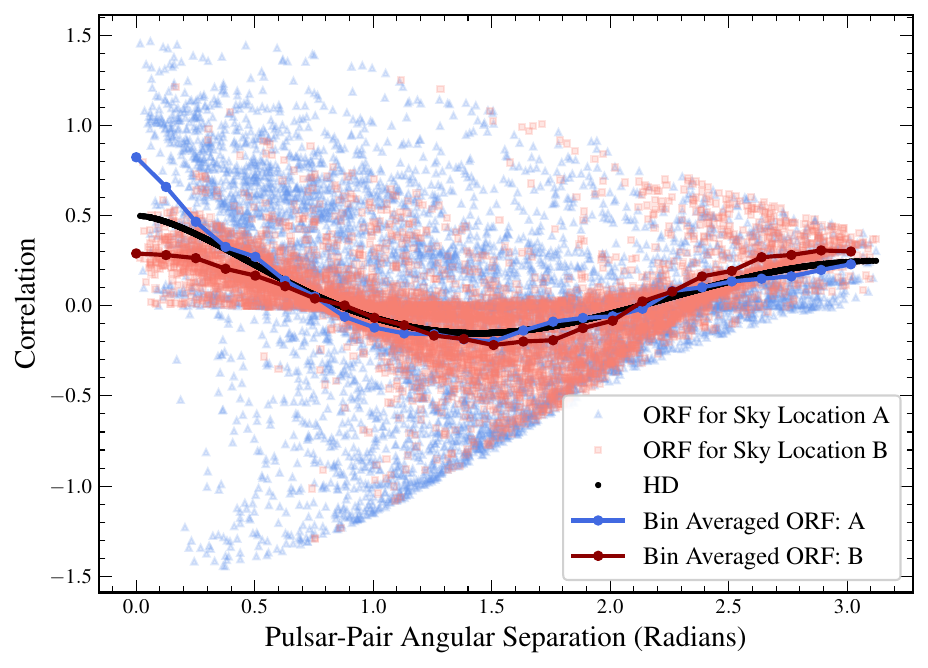}
    \caption{The ORF induced by a single point of GW emission on the sky, as would be created by an individual SMBHB. Shown are the cross correlations for pulsars from our simulated datasets (described in \autoref{sec:simdatasets}). Correlations are binned arbitrarily for ease of viewing the general structure and comparing to the expectation from an isotropic sky (HD).}
    \label{fig:pixorf}
\end{figure}

Example inter-pulsar correlations for the synthetic datasets we later use are computed via \autoref{eqn:pixorf} and shown in \autoref{fig:pixorf}. This ORF, combined with the PSD for a single spike of power at a frequency, $f'$, gives the cross-PSD of the SP model. The cross-PSD model for the entire signal is the addition of the statistically isotropic GWB model and the SP model, giving
\begin{equation}
        S_{ab}(f) = S_\mathrm{GWB}(f)\Gamma^{\rm HD}_{ab} + S_{\mathrm{binary}}(f')\aleph_{ab}(\hat{\Omega}') \delta(f -f'),
\end{equation}
where $S_\mathrm{binary}(f')$ is the PSD of a single binary with GW frequency, $f'$. $S_\mathrm{binary}(f')$ is defined as
\begin{equation}
    S_\mathrm{binary}(f') = \frac{\rho_{\rm SP}} {\Delta f}=\rho_{\rm SP} \times T_{\rm span}
\end{equation}
where $T_{\rm span}$ is the observing baseline of the array and $\rho_{\rm SP}$ is the Fourier domain auto-covariance of timing residuals ($\rho_{\rm SP}$ is derived in \autoref{sssec:rhospderiv}).

\subsection{Data Model}
We briefly summarize the PTA data model and likelihood, following more detailed treatments \citep{2017PhRvD..95d2002T, 2021arXiv210513270T}. We can write the vector of timing residuals obtained upon removing a best-fit timing solution from the observed TOAs as 
\begin{equation}
    \vec{\delta t} = \mathbf{M}\vec\epsilon + \mathbf{F}\vec{a} + \vec{n},
\end{equation}
where $\mathbf{M}$ is the $N_\mathrm{TOA}\times N_\epsilon$ timing-model design matrix, corresponding to partial derivatives of each TOA with respect to each timing model parameter; $\vec\epsilon$ is a vector of linear offsets to each parameter in the timing model; $\mathbf{F}$ is the $N_\mathrm{TOA}\times 2N_f$ Fourier design matrix, which contains a sine and cosine basis for each frequency being sampled; and $\vec{a}$ is a vector of Fourier coefficients. The sampling frequencies are $i/T$ where $T$ is the total timespan of the dataset, and $i$ is an integer that iterates frequency-resolution bins. Finally, $\vec{n}$ is Gaussian noise. With this, Gaussian residuals that now include noise processes and any signal from a GWB can be written as
\begin{equation}
\label{eqn:residsdeflikelihood}
    \vec{r} = \vec{\delta t} - \mathbf{M}\vec{\epsilon} - \mathbf{F}\vec{a} = \vec{\delta t} - \mathbf{T}\vec{b}
\end{equation}
whose likelihood is that of a multivariate Gaussian,
\begin{equation}
\label{eqn:wnmvgauss}
    p(\vec{\delta t} | \vec{\epsilon}, \vec{a}) = \mathcal{N}(\mathbf{T}\vec{b}, \mathbf{N}), 
\end{equation}
where $\mathbf{N}$ is the white noise covariance matrix of the TOAs in the dataset, and the grouped design matrices and coefficients are
\begin{equation}
    \textbf{T} = [\textbf{M F}],\,\,\, \vec{b} = \begin{bmatrix} \vec{\epsilon} \\ \vec{a}
    \end{bmatrix}.
\end{equation}

We apply a Gaussian prior on $\vec{b}$, such that $p(\vec{b}|\varphi) = \mathcal{N}(0,\mathbf{B})$, where the covariance matrix of low-rank processes, $\mathbf{B}$, is
\begin{equation}
\label{eqn:Bcoeffpriors}
    \textbf{B} = \begin{bmatrix}
        \infty & 0 \\ 0 & \varphi
    \end{bmatrix}.
\end{equation}

This covariance matrix imposes an infinite variance prior on the linear timing-model offsets to approximate an unbounded uniform prior. 
The term $\varphi$ represents the variance of the Fourier coefficients governing red noise in the data. 
As previously stated, an isotropic GWB creates a red, stochastic process that is present across pulsars in the array with HD spatial correlations. Individual pulsars also have their own low-frequency noise independent of the GWB. Thus, both must be included in $\varphi$ \cite{2013PhRvD..87j4021L, 2017PhRvD..95d2002T}, making the matrix index over both pulsars ($a, b$ indices) and the sampled frequencies ($i, j$ indices):
\begin{equation}
\label{eqn:chiorphi}
    [\varphi]_{(ai), (bj)} = \Gamma_{ab}^{\rm HD} \beta_i \delta_{ij} + \aleph_{ab} \rho_i \delta_{ij} + \kappa_{ai}\delta_{ab}\delta_{ij}, 
\end{equation}
where the first term models the underlying statistically-isotropic GWB, the second term models the excess GW power of a single binary, and the final term accounts for intrinsic (uncorrelated) red noise in each pulsar. Each term has a variance factor ($\beta$, $\rho$, $\kappa$) corresponding to the PSD of the induced timing residuals divided by the dataset timespan. 
We use a power-law model to describe $\beta$ and $\kappa$, both following the form
\begin{equation}
\label{eqn:gwbpowerlaw}
    \beta_i = \frac{A^2}{12 \pi^2 T}\left(\frac{f_i}{f_\mathrm{yr}}\right)^{-\gamma}  \mathrm{yr}^2.
\end{equation}

While the amplitude, $A$, and spectral index, $\gamma$, parameters are common across all pulsars in the array for the GWB, the intrinsic red noise is modeled with pulsar-specific versions of these parameters. We group all hyperparameters of the red noise processes into the vector $\vec\eta$, such that $\varphi = \varphi(\vec\eta)$.

\subsubsection{Data likelihood}

The hierarchical likelihood for the PTA data model is then
\begin{equation}
\label{eqn:hierlikenotmarg}
    p(\vec{\delta t} | \vec{b}, \vec\eta) \propto p(\vec{\delta t} | \vec{b}) \times p(\vec{b}|\eta)
\end{equation}
over which we can analytically marginalize the low-rank coefficients, $\vec{b}$ \cite{2013PhRvD..87j4021L, 2013MNRAS.428.1147V}, such that the marginalized likelihood is 
\begin{equation}
\label{eqn:ptalike}
    p(\vec{\delta t} | \vec\eta) \propto \frac{\exp(\frac{1}{2}\vec{\delta t}^{T}\textbf{C}^{-1}\vec{\delta t})}{\sqrt{\det(2 \pi \textbf{C})}}
\end{equation}
where $\textbf{C} = \textbf{N} + \textbf{TBT}^T$. It is also common practice to reduce the computational burden of this likelihood by neglecting inter-pulsar correlations, and instead modeling only the autocorrelations of a GWB. This common spatially-uncorrelated red noise (CURN) model is a spectral proxy for a GWB, and is easier to compute as it makes $\varphi$ diagonal, and thus trivial to invert in a likelihood evaluation.

Modeling additional deterministic signals within the framework of \autoref{eqn:ptalike} is trivial; we replace $\delta \vec{t}$ with $\vec{\delta t} - \vec{s}(\vec{\lambda})$ where $s$ is our deterministic model, in this case a GW from a SMBHB that depends on parameters $\vec{\lambda}$. This enables us to model a stochastic background as well as a discrete binary simultaneously. We adopt uniform priors for the waveform parameters, with bounds described in \autoref{tab:priortable}.

\subsubsection{Timing residual variance due to a single binary}
\label{sssec:rhospderiv}

The Fourier-domain auto-covariance, $\rho_{\rm SP}$, of timing residuals induced by a GW from a single SMBHB is defined in terms of binary parameters. Beginning with \autoref{eqn:circcwresid}, one takes the variance of $s_{A}$ (recalling that $\langle s_{A} \rangle = 0$):
\begin{equation} \label{eq:spluscross_var}
    \langle s_{+}^2 \rangle = \langle s_{\times}^2 \rangle =
        \frac{4}{5}\left(\frac{\mathcal{M}^{5/3}}{d_L \omega_0^{1/3}}\right)^2 .
\end{equation}

The cross-covariance of timing residuals is
\begin{equation}
\label{eqn:rhointro}
    \langle s_a s_b \rangle = F^{+}_a F^{+}_b \langle s_{+}^2 \rangle + F^{\times}_a F^{\times}_b \langle s_{\times}^2 \rangle,
\end{equation}
into which we substitute \autoref{eq:spluscross_var} to get
\begin{equation}
    \langle s_a s_b \rangle = \frac{4}{5}\left(\frac{\mathcal{M}^{5/3}}{d_L \omega_0^{1/3}}\right)^2 \left[F_{+}^2 + F_{\times}^2\right],
\end{equation}
where the term in squared brackets can be recast in the form of the single-pixel ORF from \autoref{eqn:pixorf}:
\begin{align}
    \langle s_a s_b \rangle &= \frac{4}{5}\left(\frac{\mathcal{M}^{5/3}}{d_L \omega_0^{1/3}}\right)^2 \frac{2}{3} \aleph_{ab}(\hat{\Omega}') \nonumber\\
    &= \frac{8}{15} \left(\frac{\mathcal{M}^{5/3}}{d_L \omega_0^{1/3}}\right)^2 \aleph_{ab}(\hat{\Omega}') \nonumber\\
    &= \rho_\mathrm{SP}\times\aleph_{ab}(\hat{\Omega}')
\end{align}

\subsubsection{\textit{Spike pixel} model implementation}

We search over the sky position of the GW source through the ORF, $\aleph_{ab}(\hat\Omega')$, remembering that the unit vector pointing to the source is $-\hat\Omega'$. Our  prior distributions on source position are uniform in $\cos\theta \in [-1, 1]$ and $\phi \in [0, 2\pi]$. When searching, these parameters are cast to a single pixel determined by the resolution of our equal-area pixelization constructed via \texttt{HEALPix}. We use $N_\mathrm{side}=32$ which forms a skymap with $N_\mathrm{pixels}=12N_\mathrm{side}^2=12288$. 

This radiometer-type search is limited to a single GW frequency bin, which is also data-determined during the search. This frequency bin has an associated timing-residual variance parameter $\rho_\mathrm{SP}$, with a prior such that $\log_{10}(\rho_\mathrm{SP}/\mathrm{sec^2)} \sim U[-18, -8]$, as is common in the literature \cite{2023ApJ...951L...8A, 2020ApJ...905L..34A}. While the frequency bin index ($f_\mathrm{SP}$) is a discrete parameter, we search continuously with a uniform prior in the log of the frequency-bin index, which is exponentiated and then cast down to the lowest integer. This prior covers the range $\in [\log_{10}(0.1), \log_{10}(N_\mathrm{components}-0.01)]$. $N_\mathrm{components}$ is the number of components used to model the GWB. This log-uniform prior is used to give the SP model's frequency parameter consistency with the scale-invariant prior on GW frequency used in deterministic-templated searches.

The parameters for intrinsic pulsar red noise and the underlying isotropic GWB are the hyperparameters governing their power-law spectra. These are modeled on a grid of frequencies of size $N_{\mathrm{components}}=T_\mathrm{span}/(1\,\mathrm{year})$, rounded down to the nearest integer, ensuring that the maximum frequency modeled stays the same as we extend our datasets. 

To perform the Bayesian analyses of our simulated PTA datasets, we use MCMC techniques \citep{1970Bimka..57...97H} to draw random samples from the joint posterior probability distribution of signal and noise parameters. Specifically, we use the \texttt{QuickCW} software package from \citet{2022PhRvD.105l2003B} to explore the deterministic-template model for the full CW signal. While this package can efficiently search over the many parameters (8 + 2$\mathrm{N_{pulsar}}$) included in a full signal continuous wave model (hereafter referred to as $\mathrm{CW_{FS}}$), it is not currently able to include HD correlations in its model for the GWB. Hence, the implementation used in this paper uses a CURN model for the background. We do not ameliorate this issue using reweighting \citep{2023PhRvD.107h4045H} due to poor efficiency. For more details see \autoref{subsec:disc-modimp}.
When we search with the simpler Earth-term-only CW model (hereafter referred to as $\mathrm{CW_{ET}}$), we use \texttt{ENTERPRISE Extensions}, an add-on package to \texttt{ENTERPRISE} \cite{2019ascl.soft12015E, enterprise}. This software enables us to build our signal model for a CW with HD spatial correlations included for the GWB. For sampling the $\mathrm{CW_{ET}}$ and our SP anisotropy model, we use a bespoke parallel-tempering MCMC sampler for PTA searches, \texttt{PTMCMCSampler} \cite{justin_ellis_2017_1037579}.

\begin{table*}
\caption{\label{tab:priortable}Parameter priors used across all \textit{Spike Pixel} (SP), Earth Term continuous wave ($\mathrm{CW_{ET}}$), and full signal continuous wave ($\mathrm{CW_{FS}}$) analyses. From top to bottom, the parameters are grouped in sections according to those that are included across all three analyses, those that are used only in the SP analysis, those that are used in the $\mathrm{CW_{ET}}$ and $\mathrm{CW_{FS}}$ analyses, and those that are used only in the $\mathrm{CW_{FS}}$ analysis.}
\begin{ruledtabular}
\begin{tabular}{ccccc}
Parameter & SP & $\mathrm{CW_{ET}}$ &
$\mathrm{CW_{FS}}$ & Prior \\
\hline
$A_{\rm{RN}}$ & \cmark & \cmark & \cmark & Uniform($-$18, $-$11) \\
$\gamma_{\rm{RN}}$ & \cmark & \cmark & \cmark & Uniform(0, 7) \\
$A_{\rm{GWB}}$ & \cmark & \cmark & \cmark & Uniform($-$18, $-$11) \\
$\gamma_{\rm{GWB}}$ & \cmark & \cmark & \cmark & Uniform(0, 7) \\
$\cos\theta$ & \cmark & \cmark & \cmark & Uniform($-$1, 1) \\
$\phi$ & \cmark & \cmark & \cmark & Uniform(0, 2$\pi$) \\
\hline
$\log_{10}(\rho_{\mathrm{SP}}/\mathrm{sec^2})$ & \cmark & \xmark & \xmark & Uniform($-$18, $-$8) \\ 
$f_\mathrm{SP}$ & \cmark & \xmark & \xmark & Uniform($\log_{10}(.1)$, $\log_{10}(N_\mathrm{years}-0.01)$) \\
\hline
$\log_{10}(f_{\rm{GW}}/\rm{Hz})$ & \xmark & \cmark & \cmark & Uniform($-$9, $-$7.5) \\
$\log_{10}h_0$ & \xmark & \cmark & \cmark & Uniform($-$18, $-$11) \\
$\log_{10}(\mathcal{M}/M_{\odot})$ & \xmark & \cmark & \cmark & Uniform(7, 10) \\
$\cos\iota$ & \xmark & \cmark & \cmark & Uniform($-$1, 1) \\
$\psi$ & \xmark & \cmark & \cmark & Uniform(0, $\pi$) \\
$\Phi_0$ & \xmark & \cmark & \cmark & Uniform(0, 2$\pi$) \\
\hline
$L_p$ & \xmark & \xmark & \cmark & Normal($L_p$, $\sigma_{L_p}$) \\
$\Phi_p$ & \xmark & \xmark & \cmark & Uniform(0, 2$\pi$) \\
\end{tabular}
\end{ruledtabular}
\end{table*}

\subsection{Detection Statistics}

An important question is which approach to modeling an emerging SMBHB GW signal will show the greatest growth in a detection statistic. We address this with the Savage-Dickey density ratio (or Savage-Dickey Bayes Factor; SDBF), an exact Bayesian evidence ratio between two models: one which includes an individual SMBHB GW signal along with noise and an underlying GWB (model $M_{\rm CW+GWB}$), versus one in which the individual signal is absent (model $M_{\rm GWB}$) \cite{dickey1971weighted, 2021arXiv210513270T}. The SDBF is appropriate since all models for the inclusion of a binary signal are nested, wherein the likelihood under $M_{CW+GWB}$ at the lowest amplitude of the binary signal approximates the likelihood of $M_{\rm GWB}$, i.e.,
\begin{multline}
    p(\vec{\delta t}|\vec\theta; M_{\rm GWB}) = \\ p(\vec{\delta t}|\{h_0, \rho_{\rm SP} \} = 0, \vec\theta; M_{\rm CW+GWB})
\end{multline}

The Bayes factor of interest then becomes the ratio of prior to posterior at the lowest binary signal amplitude:
\begin{equation}
\label{eqn:sdbf_rawmath}
    \mathcal{B}_{\rm(CW+GWB)/GWB} = \frac{\pi(\{h_0, \rho_{\rm SP} \} = 0)}{p(\{h_0, \rho_{\rm SP} \} = 0|\vec{\delta t}; M_{\rm CW+GWB})}.
\end{equation}

The SDBF is exact in its relation to the ratio of Bayesian evidences, but becomes approximate through its calculation via Monte Carlo samples drawn from the posterior. This is ultimately its limitation, since this relies on there being some MCMC exploration of the parameter space in which the binary amplitude is zero (or close to the minimum of the prior boundary). As the support for the binary grows, the posterior on SMBHB parameters becomes more constrained, and it is less likely for the sampler to explore such low-probability regions of parameter space. When this happens (which we shall encounter), the SDBF cannot be calculated due to inadequate sampling coverage; we can then only place a lower limit. 
One can numerically estimate the denominator in \autoref{eqn:sdbf_rawmath} as,
\begin{equation}
    p(A_{\rm CW}=0|\delta\vec{t}; M_{\rm CW+GWB}) = \frac{n_\mathrm{count}}{N_\mathrm{samples}} \times \frac{1}{w_\mathrm{bin}},
\end{equation}
where $n_\mathrm{count}$ is the number of samples that fall in a low amplitude region of the posterior with width $w_\mathrm{bin}$ and an MCMC chain length of $N_\mathrm{samples}$. We average our SDBF estimates over a range of posterior histogram bin widths and tolerance sample counts (if $n_\mathrm{count} < n_\mathrm{tolerance}$, we skip the calculation due to inadequate sample coverage).\footnote{$N_\mathrm{tol}$=[1, 50, 75, 100, 150, 200] and $N_\mathrm{bin}$=[20, 50, 75, 100, 150, 200]. We include an $N_\mathrm{tol}$ of 1 for datasets in which support for the signal is overwhelmingly strong, leading to very few samples in the region of interest.} 
If there are no samples in the range examined, we determine a lower limit by assuming there is precisely one posterior sample in the lowest amplitude bin, such that $\mathcal{B}_\mathrm{low} = N_\mathrm{samples}/N_\mathrm{bins}$. If the SDBF calculations fail in half of the noise realizations of a given time slice,
we do not report the Bayes factors. 
As a guide to quantifying the strength of the CW signal in our datasets, we use the logarithm of the ratio of likelihoods, $\ln\Lambda$, between $M_{\rm CW+GWB}$ and $M_{\rm GWB}$, computed with injected parameter values. 
This gives
\begin{equation}
\label{eqn:likeratio_hlamb}
    \ln \Lambda = \ln \mathcal{L}_{\rm CW+GWB} - \ln \mathcal{L}_{\rm GWB} = (\vec{\delta t}| s(\vec{\lambda})) - \frac{1}{2} (s(\vec{\lambda})|s(\vec{\lambda})),
\end{equation}
where the notation $(\textbf{x}|\textbf{y}) = \textbf{x}^T \textbf{C}^{-1} \textbf{y}$ corresponds to a noise-weighted inner product with a covariance matrix that includes all intrinsic pulsar noise and stochastic GW signals.

This log-likelihood ratio can be converted into a signal-to-noise ratio (S/N) such that $\mathrm{(S/N)}_{\Lambda} = \sqrt{2 \ln \Lambda}$, where $\ln \Lambda$ is calculated with the injected parameters. This (S/N) can be evaluated dataset realization-by-realization, and for all our models of the SMBHB signal (including the SP model). It gives a measure of the (S/N) achieved by our different SMBHB signal models in each dataset. We can also use \autoref{eqn:likeratio_hlamb} to form the optimal (S/N), given by setting the GW signal template to the true signal, and averaging over noise realizations that have zero match (on average) with the signal template; $\mathrm{(S/N)_{opt}} = (s|s)^{1/2}$. The optimal (S/N) gives a measure of the average maximum (S/N) that one could achieve through matched filtering. 

\section{Simulated Datasets}
\label{sec:simdatasets}
\begin{figure}[!ht]
    \centering
    \includegraphics[width=0.5\textwidth]{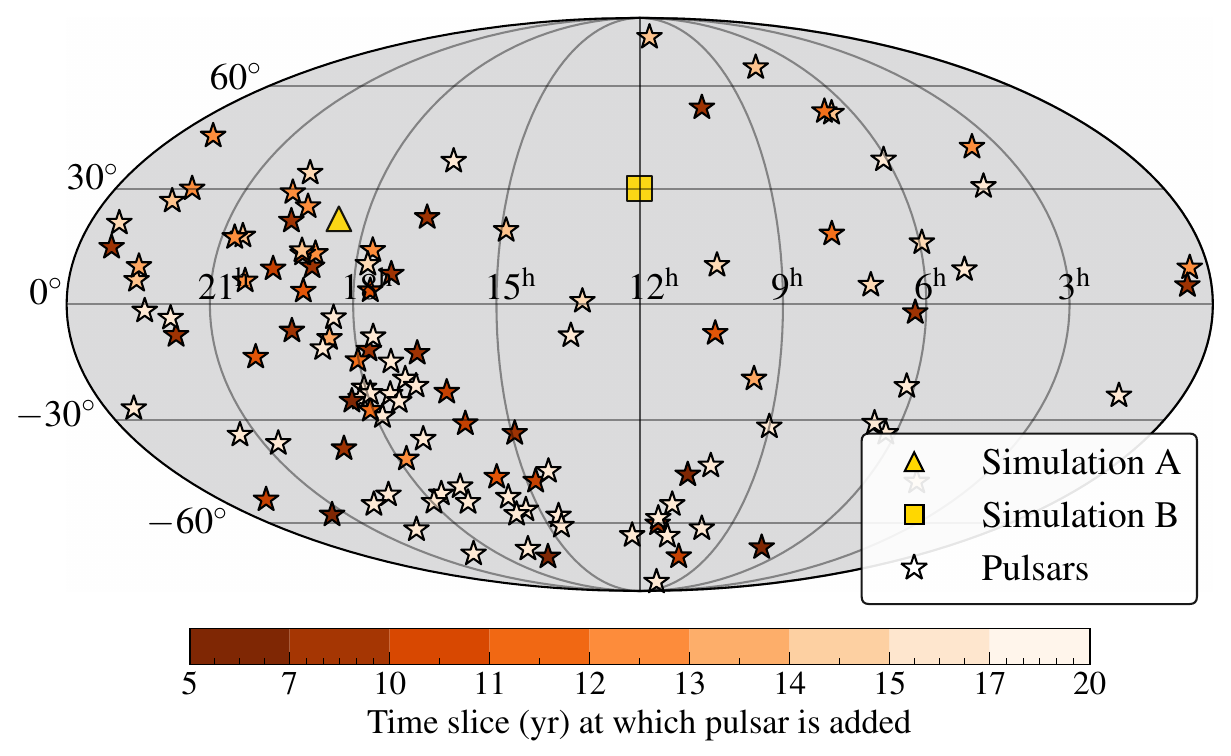}
    \caption{Sky map of all pulsar positions in our simulated array, along with the two sky locations into which we inject a CW signal. The pulsars are colored according to the time slice in which they are added to the dataset, with the earliest-added (and consequently longest-timed) pulsars represented by darker orange stars, while pulsars added to the array more recently are represented by progressively lighter orange stars. The injected sky locations for Simulations A and B are shown with a yellow triangle and square, respectively. By the 20-year dataset, the array includes all 116 pulsars.}
    \label{fig:skymap}
\end{figure}

\begin{table}
\caption{\label{tab:simtable} Injected CW parameters for each simulation set. All injections have a chirp mass of $\mathcal{M}=10^9 M_{\odot}$, face-on binary inclination angle $\iota = 0$, GW polarization angle $\psi = \pi/4$, and initial binary phase $\Phi_0 = \pi/4$. Each source's $d_L$ is chosen so that the CW signal reaches (S/N)$_{\rm opt}$~$\sim$~20 by the final 22-year time slice.}
\begin{ruledtabular}
\begin{tabular}{ccccc}
\multirow{2}{*}{Simulation} & \multirow{2}{*}{$\theta$ (rad)} &
\multirow{2}{*}{$\phi$ (rad)} & \multicolumn{2}{c}{$d_L$ (Mpc)} \\
& & & 5 nHz & 20 nHz \\
\hline
A & 1.19 & 4.87 & 33.75 & 124.5 \\
B & 1.05 & 3.14 & 19.42 & 69.75 \\
\end{tabular}
\end{ruledtabular}
\end{table}

\begin{table}
\caption{\label{tab:slicetable} Number of pulsars included in each time slice of our simulated IPTA-DR3-inspired dataset.}
\begin{tabular}{c | c}
\hline
Time slice (yr) & Number of pulsars \\
\hline
\hline
5 & 5 \\
7 & 20 \\
10 & 27 \\
11 & 32 \\
12 & 36 \\
13 & 47 \\
14 & 50 \\
15 & 58 \\
17 & 66 \\
20 & 116 \\
22 & 116 \\
\hline
\end{tabular}
\end{table}

Our simulated, realistic PTA datasets are inspired by the forthcoming IPTA-DR3. Following the framework in \citet{astro4cast} and \citet{2024ApJ...976..129P}, we start with the 45 pulsars in the NANOGrav 12.5-year dataset \citep{2021ApJS..252....4A}, using the real observing schedule, TOA uncertainties, and timing models measured in this dataset, and then extend these observations to a 20-year baseline. To reach 20 years, additional observational data are generated with new TOA uncertainties and a new cadence by drawing from distributions based on the dataset's last year of observations \citep{astro4cast}. 

Following these 45 pulsars, we add on 23 pulsars from the NANOGrav 15-year dataset \citep{2023ApJ...951L...9A} (giving a total of 68 NANOGrav pulsars in our simulated PTA), 14 pulsars from PPTA DR3 \citep{2023arXiv230616230Z}, 3 pulsars from EPTA+InPTA DR2new+ \citep{2023arXiv230616224A}, and 31 pulsars from MPTA DR1 \citep{2023MNRAS.519.3976M}, following the simulation setup outlined in \cite{2024ApJ...976..129P}. These 71 pulsars not in the NANOGrav 12.5-year dataset retain their true sky positions but borrow timing model parameters from an ``average” template pulsar. This template pulsar is chosen at random from one of J0931-1902, J1453+1902, J1832-0836, or J1911+1347, all of which were added to the array between the NANOGrav 11-year and 12.5-year data sets and are not among the current most sensitive pulsars. For any pulsar timed by multiple PTAs, we use its NANOGrav-observed baseline if it is timed by NANOGrav; otherwise, we take the pulsar to be part of the array in which its baseline is longest. The pulsars are timed over these real baselines, with more observations added to reach a 20-year NANOGrav baseline. We note that while the dataset's baseline spans 20 years for the NANOGrav sub-array of pulsars, the baseline of the entire 116-pulsar array is $\sim$ 22 years. Each pulsar's observations are taken every two weeks, and the TOA uncertainties come from the whitened RMS values reported in the pulsar's respective dataset paper. All pulsar distances are the same as those used in \citet{2024ApJ...976..129P}.

In \autoref{fig:skymap} we show the positions of all 116 pulsars in the array on a sky map, along with the two sky locations into which we inject a CW signal. The sky location labeled Simulation A (yellow triangle) contains a CW signal injected at a location that is favorable to detection and localization, ($\theta$, $\phi$) = (1.19, 4.87) radians or $(\alpha,\delta) =$ (18h36m, $+22^{\circ}$), while the sky location labeled Simulation B (yellow square) contains a CW signal in a less favorable position, ($\theta$, $\phi$) = (1.05, 3.14) radians or $(\alpha, \delta) =$ (12h, $+30^{\circ}$).\footnote{Favorable and unfavorable qualifiers derive from Figs 1 \& 5 of \citet{2024ApJ...976..129P} and are based on the proximity of pulsars to the source.} For each sky location, we generate two sets of simulations, each with a different GW frequency. The first set contains a signal with $f_{\rm{GW}}$ = 5 nHz, which does not have significant evolution between the Earth term and pulsar terms, while the second set contains a signal with $f_{\rm{GW}}$~=~20 nHz, producing a more pronounced evolution between the two terms. Each CW source has a chirp mass $\mathcal{M} = 10^9$~M$_{\odot}$, and we adjust the luminosity distance of each source to fix (S/N)$_{\rm opt} \sim$ 20. All injections have a face-on inclination angle $\iota=0$, GW polarization angle $\Psi = \pi/4$, and initial binary phase $\Phi_0 = \pi/4$. A summary of the four sets of simulations and their respective injected CW parameters can be found in \autoref{tab:simtable}.

Each of the four simulation sets (A 5 nHz, A 20 nHz, B 5 nHz, B 20 nHz) are created with ten different stochastic-process realizations, in which the time-series of white noise, intrinsic pulsar red noise, and GWB all vary from realization to realization. We use the reported uncertainties on TOAs and do not apply any multiplicative correction factor (EFAC=1). Each pulsar's intrinsic red noise parameters are injected at the values measured in their respective dataset papers. The GWB is injected as a spatially-correlated red noise process with a power law spectral index $\gamma_{\mathrm{GWB}} = 13/3$ corresponding to a background produced by a population of circularly-inspiraling SMBHBs \citep{Phinney01} and amplitude $A_{\mathrm{GWB}}$ = 2.4~$\times$~$10^{-15}$ as reported in the NANOGrav 15-year dataset \citep{2023ApJ...951L...8A}. 

Finally, to track the time-evolving efficacy of our search procedures, we slice each dataset into sub-datasets with 11 different timespans. We do this by taking the first TOA of a given dataset and simply truncating the dataset at the desired length of time. We also require that each pulsar has $\geq$ 3 years of timing data before being added to the array, mirroring the requirement used in real PTA datasets; therefore, if at any time slice there exists a pulsar with a timing baseline $<$ 3 years, it is excluded from that slice. Our time slices have timespans of 5, 7, 10, 11, 12, 13, 14, 15, 17, and 20 years, and the final 22-year slice is the full, original dataset. These intervals were chosen by virtue of their $\mathrm{(S/N)_{\Lambda}}$ ranges. The 10-15 year slices have an average $\mathrm{(S/N)_{\Lambda}}$ range of 6-12, where individual binary detection is expected to occur \citep{ 2025arXiv250216016G, 2022ApJ...941..119B, 2018MNRAS.477..964K}. We do not expect there to be drastic changes in the binary characterization at very early slices (low $\mathrm{(S/N)_{\Lambda}}$) or late slices (high $\mathrm{(S/N)_\Lambda
}$), hence their coarser sampling.

The various slices can be seen more clearly in \autoref{fig:skymap}, where the pulsars are colored by the time slice at which they are added to the dataset. The darkest orange stars on the sky map represent the earliest-added 5-year time slice pulsars, while increasingly lighter stars represent pulsars added at later time slices. Because the darkest, earliest-added pulsars appear in every subsequent dataset, they are also the longest-timed; conversely, the lighted, latest-added pulsars are the shortest-timed. We note that the color bar shading in \autoref{fig:skymap} stops at the 20-year time slice, as this is the dataset at which all 116 pulsars have at least three years of timing data and have been added to the array. The number of pulsars in each respective time-sliced dataset can be found in \autoref{tab:slicetable}. 

Our grand total of simulations amounts to 2 sky locations $\times$ 2 GW frequencies $\times$ 10 noise realizations $\times$ 11 time slices = 440 datasets. Each was searched using the three approaches previously described: the SP anisotropy model, an Earth-term-only waveform template search ($\mathrm{CW_{ET}}$), and a full signal template search ($\mathrm{CW_{FS}}$).

\section{\label{sec:Results} Results}

\subsection{Comparing the Detection Efficacy of Models}
\label{sec:detectstats}

\begin{figure}
    \centering
    \includegraphics[width=0.49\textwidth]{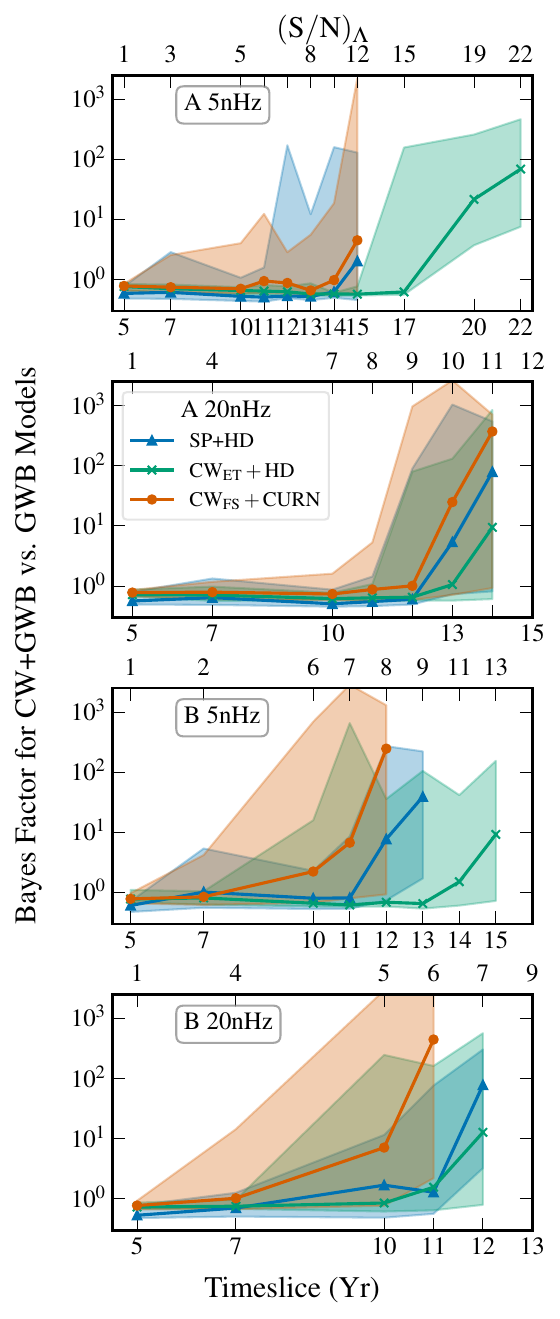}
    \caption{We show the median and full range of the Savage-Dickey Bayes factor for each binary search technique with respect to the dataset timespan. The corresponding $\mathrm{(S/N)_{\Lambda}}$ is plotted on the top $x$-axis. The full signal model $\mathrm{CW_{FS}}$  (orange circles) recovers higher BFs in comparison to the coarse SP anisotropy model (blue triangles) and the $\mathrm{CW_{ET}}$ (green $\times$s). When there is insufficient MCMC sampling coverage of the CW amplitude's posterior tail, an upper limit is reported for that realization. We do not report BFs if half the realizations require such upper limits. 
    }
    \label{fig:sdbf_acrossmodels}
\end{figure}

We first compare the Bayes factors in favor of each model with respect to just noise (and an isotropic GWB), to score their ability to identify the emerging binary signal in our data. In \autoref{fig:sdbf_acrossmodels}, we show the realization-median SDBF for each model across time slices, as well as the full span of values over realizations. The preference for a binary signal eventually exceeds the relatively weak signal assumption that the SDBF calculation relies upon. Therefore, we do not report BFs when half of the realizations  have inadequate sampling coverage for BF calculation. 

There are significant variations across binary injections and realizations, with some models recovering a spread of BFs spanning 1-3 orders of magnitude as the data becomes informative. Ranking the performance of the examined models shows that $\mathrm{CW_{FS}}$ tends to have the highest BFs, followed by SP and $\mathrm{CW_{ET}}$. However, we note that there are some time slices in particular realizations where this model hierarchy is different. The hierarchy for detection across injections is clear, with B 20 nHz being detected earliest, followed by B 5 nHz, A 20 nHz, and finally A 5 nHz. The differences in performance between the four injections are discussed in Paper II \citep{2025arXiv251001316P}. 

The $\mathrm{CW_{FS}}$ model attains a median $\mathrm{BF} > 10$ when $\mathrm{(S/N)_\Lambda} \sim6-15$ across the four binary injections. In this regime of BF growth, the median BF of the SP model is, in general, 1-2 orders of magnitude less than that of the full signal template. This translates to SP reaching the BF $> 10$ threshold about one year (or an increase in $\mathrm{(S/N)_\Lambda}$ of $\sim 1$) after $\mathrm{CW_{FS}}$. However, this lag in BF growth is less pronounced for some injections.
The $\mathrm{CW_{ET}}$ model has comparable BF growth to the other two models in the 20 nHz injections, but it struggles to detect the low-frequency binaries. This is consistent with behavior observed in \citet{2024arXiv240721105F}, where, due to the minimal frequency evolution between the Earth and pulsar terms, the pulsar terms act as interfering noise sources that confuse the $\mathrm{CW_{ET}}$ model. For the 5 nHz binaries, the $\mathrm{CW_{ET}}$ model attains a median BF $> 10$ at $16 < \mathrm{(S/N)}_{\Lambda} < 19$. This is in contrast to the 20 nHz case, in which $\mathrm{CW_{ET}}$ reaches the same BF threshold considerably earlier at  $7 < \mathrm{(S/ N)_{\Lambda}} < 12$ (5 years sooner in both sky locations).

\begin{figure}
    \centering
    \includegraphics[width=0.49\textwidth]{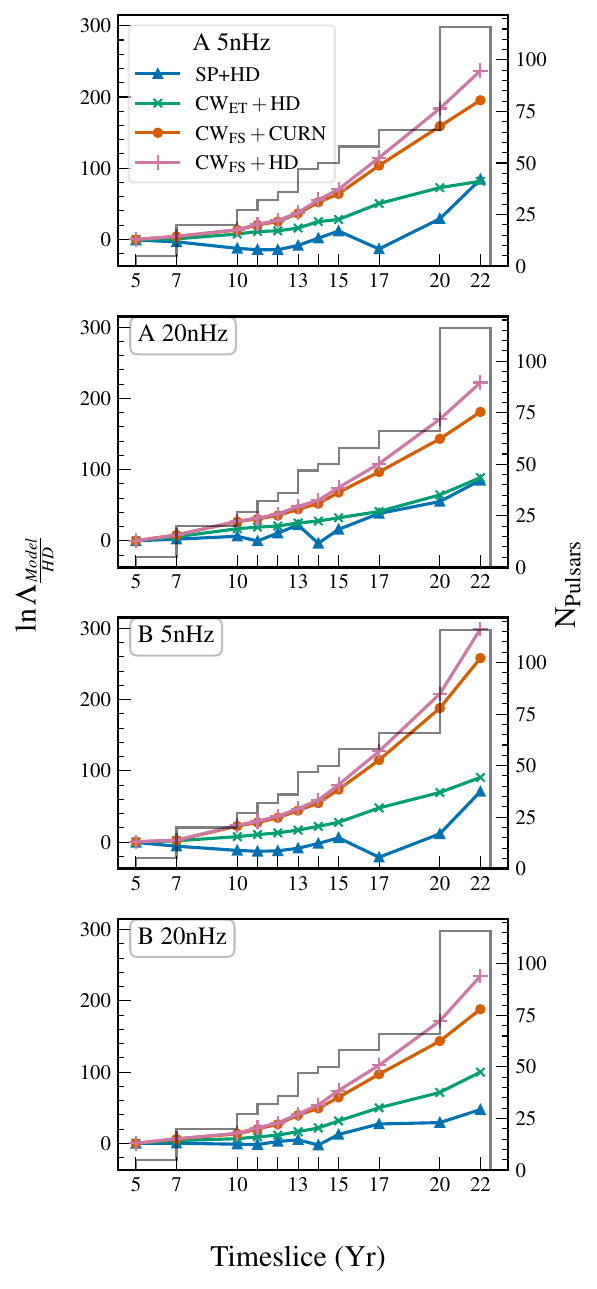}
    \caption{Since the Savage-Dickey Bayes factor becomes unreliable as the binary signal becomes stronger in the data, we turn to the ratio of log-likelihoods for each model to a HD GWB. Each log-likelihood was calculated using the injected values for each model respectively. Plotted are the medians over 100 noise realizations of the datasets, which include the ten used in the Bayesian analyses. Here we can observe the marked advantage that the full signal template model (orange dots) has over other models. We also show the true model (pink +)---which we did not apply to the dataset---to demonstrate how even the $\mathrm{CW_{FS}}$ model is impaired by misspecification by relying on a CURN model over HD for the GWB. The $\mathrm{CW_{ET}}$ model displays a higher likelihood ratio than SP, contrary to the measured BFs in \autoref{fig:sdbf_acrossmodels}. In the background, a histogram displays the number of pulsars at each time slice of the dataset.
   }
    \label{fig:lnlikeratio_acrossmodels}
\end{figure}

As we have seen, the Savage-Dickey approach to Bayes factor computation is practically limited to weak-to-moderate signals due to the need for adequate MCMC sampling coverage of the CW amplitude's posterior tail. We therefore also use the injected values to compute the ratio of each model's likelihood with respect to only stochastic GWB and noise processes to assess model support. Unlike the rest of our analyses, here we synthesize 100 realizations of each binary injection scenario, with the median $\ln\Lambda$ shown across time slices in \autoref{fig:lnlikeratio_acrossmodels}. While the ten datasets analyzed using MCMC methods are a subset of the 100 used for likelihood ratio calculations, they show similar behavior. Also shown is a model that we did not include in our Bayesian analyses due to its computational demand: $\mathrm{CW_{FS}}$ with HD spatial correlations for the GWB. The difference between the two $\mathrm{CW_{FS}}$ models is only apparent in later time slices, revealing that the CURN approach, while an incorrect prescription, can be an adequate proxy description of the GWB in early signal emergence.

The hierarchy of model performance is similar to the Bayes factor analysis. The key difference is that the SP model has a likelihood ratio less than, or comparable to, the $\mathrm{CW_{ET}}$ model for most injections and time slices. The $\mathrm{CW_{ET}}$ model having a higher likelihood ratio than SP is not unexpected, however. Since we use the injected values, the likelihood ratios denote the best possible performance of each model, probing their true aptitude for identifying the binary's signal. Thus, models that are a poorer match to the injected signal have lower likelihood ratios. $\rm{CW_{FS}+HD}$ has the highest likelihood ratios necessarily since it exactly matches what was injected into our datasets. The SP model's coarse approach similarly returns the lowest ratios.

$\mathrm{CW_{ET}}$ is a more exact signal prescription than SP. It falters in the Bayesian analysis due to model misspecification, that isn't present in the likelihood ratios due to our use of the injected values. The interference effects from the pulsar term inhibit $\mathrm{CW_{ET}}$'s ability to recognize the CW signal in the 5 nHz cases, as previously stated. The pulsar term is implicitly accounted for in the SP model via its ORF (see \autoref{eqn:pixorf}, specifically the autocorrelations), enabling its superior performance in low-frequency injections. In the 20 nHz cases, the models perform comparably, returning median BFs of the same magnitude, however, SP median BFs are usually higher. We suspect this to be the result of the simplicity of the SP model, which contains four fewer parameters than $\mathrm{CW_{ET}}$ and is agnostic to the binary phase. SP can leverage these few parameters to describe the data well during the initial stages of an emerging signal, despite being an approximate prescription. However, SP cannot surpass $\mathrm{CW_{FS}}$, since $\mathrm{CW_{FS}}$ is the true description of the binary signal. We expect that the BFs of $\mathrm{CW_{ET}}$ would eventually surpass those of SP.

\subsection{Key Parameter Estimation for EM Follow-up}
\label{subsec:res-paramest}

\begin{figure}[!ht]
    \centering
    \includegraphics[width=0.5\textwidth]{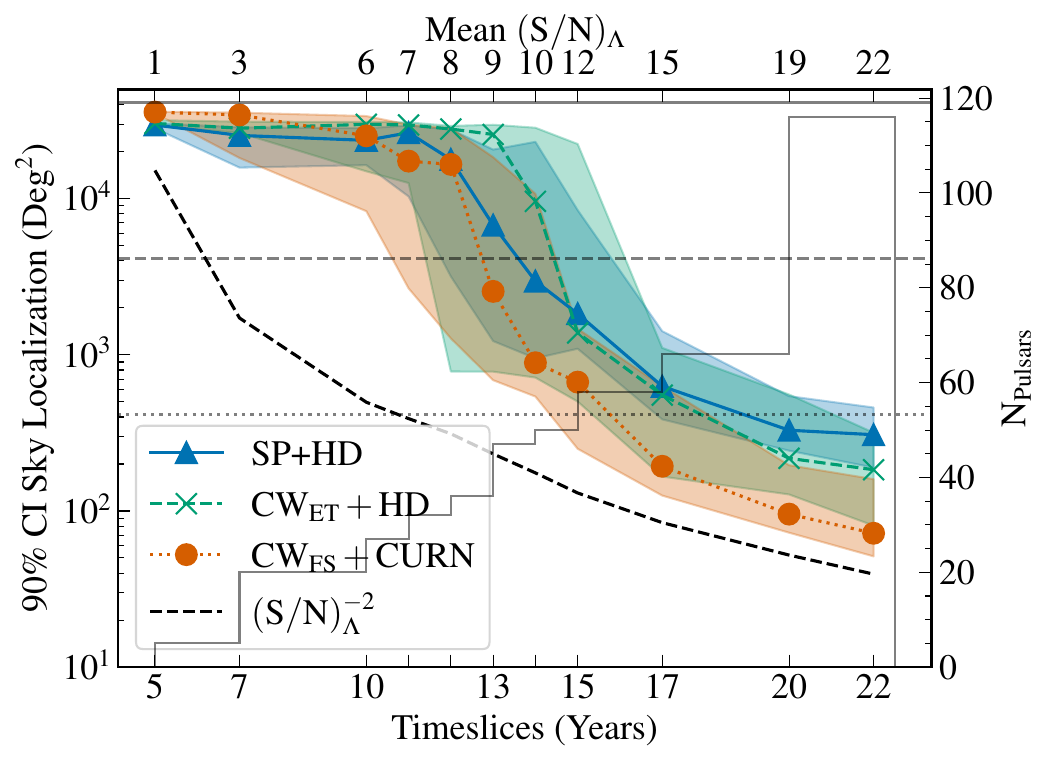}
    \caption{The colored bands and markers represent the median (over injections and realizations) 90\% credible sky localization region, while the shaded regions enclose the median range of these values. Also shown is a dashed black curve representing a $\mathrm{(S/N)}^{-2}$ relationship expected for sky localization \citep{SV2010, Taylor2016}. The grey solid, dashed, and dotted horizontal lines denote 100\%, 50\%, and 10\% of the sky, respectively. The $\mathrm{CW_{FS}}$ model returns the strongest constraints on the sky location and does so earlier than the other models. SP is the second to do so, but is surpassed by the $\mathrm{CW_{ET}}$ model at the 15-year time slice. All models narrow the source's location to $\leq$ 10\% of the sky by the final time slice, when $\mathrm{(S/N)_{\Lambda}}\sim22$.}
    \label{fig:slconstr}
\end{figure}

We first compare the precision of binary localization between the models, shown in \autoref{fig:slconstr}. We average over the four binary injections and all respective noise realizations to show how each model's constraint evolves.
Each model has a similar evolution in its localization, consisting of three distinct phases. Initially the data is uninformative, returning prior dominated constraints. There is a transition when the signal is strong enough for each respective model to hone in on the GW source's position. This stage is marked by a precipitous increase in localization precision, occurring in the 10-15 year timeslice range. The timing of this transition and how informative the data becomes in determining the sky location is dependent on the model. Finally, the models asymptote toward the expected $(\mathrm{S/N})^{-2}$ relationship expected for sky localization \citep{SV2010, Taylor2016}.

$\mathrm{CW_{FS}}$, being the most accurate prescription for the signal, necessarily restricts the GW source's position the most and earliest. Despite its coarse description of the signal, SP is able to return meaningful sky localizations earlier than $\mathrm{CW_{ET}}$, a more accurate description of the true signal.
However, the precision of the $\mathrm{CW_{ET}}$ model's sky localization eventually exceeds the SP model in later time slices. This begins at injection- and realization-averaged $\mathrm{(S/N)_\Lambda} \sim 12$, as seen in \autoref{fig:slconstr}. This matches our expectations based on the likelihood ratios found in \autoref{fig:lnlikeratio_acrossmodels}, wherein the more accurate signal prescription used in $\mathrm{CW_{ET}}$ finds greater model support than SP. 

\begin{figure*}
    \centering
    \includegraphics[width=\textwidth]{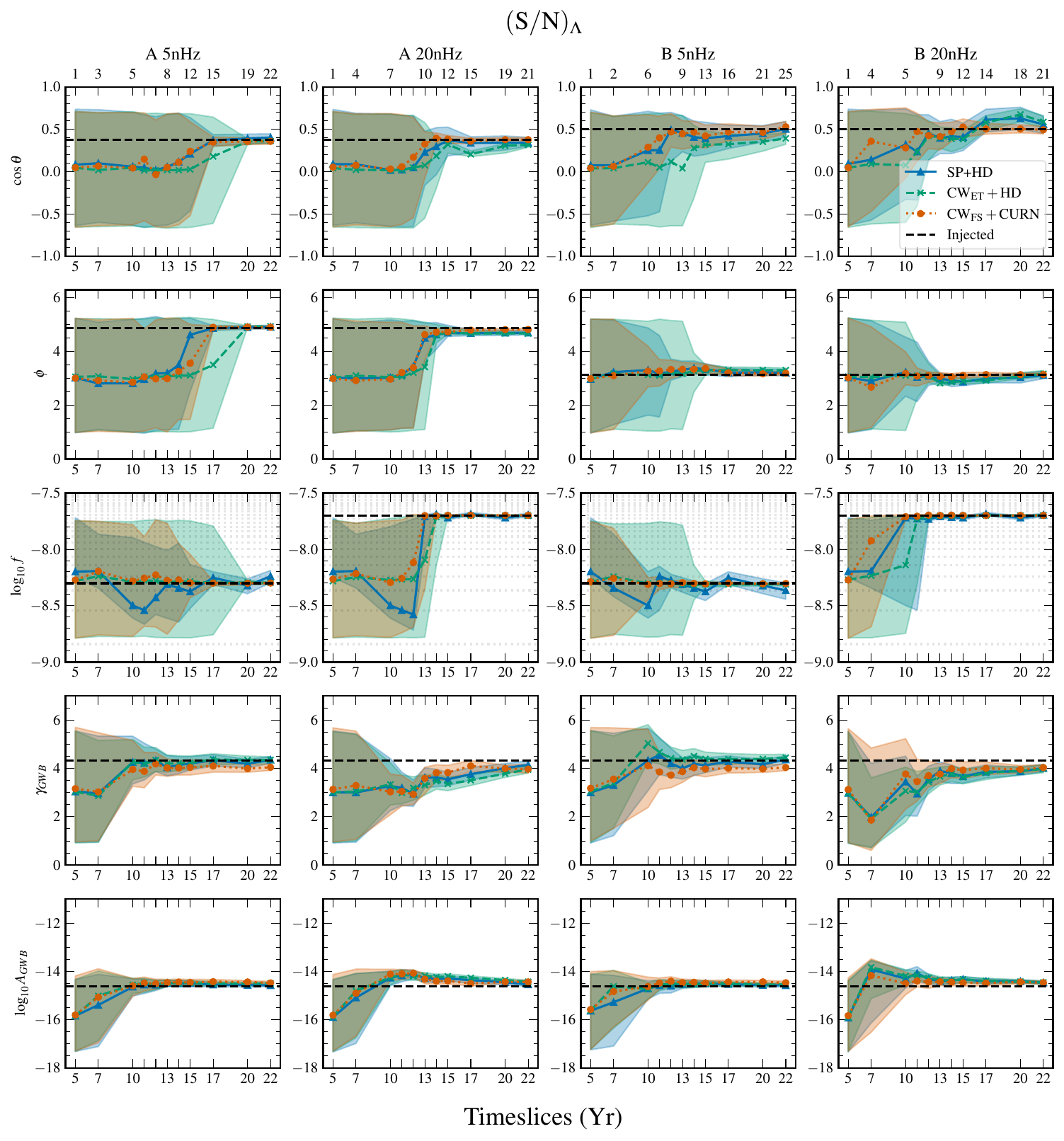}
    \caption{We show the 68\% credible intervals (shaded) and medians (lines) as a function of time slice and $\mathrm{(S/N)_\Lambda}$ for all three models across five shared parameters. The data shown are the median over the 10 realizations for each of the four binary injection suites. In each panel, the true injected value is marked by a horizontal dashed black line. Frequency bins are marked by the grey dotted lines. Model lines, markers, and colors are consistent with \autoref{fig:sdbf_acrossmodels}, \autoref{fig:lnlikeratio_acrossmodels}, and \autoref{fig:slconstr}. Of note are the biases incurred by the models in GWB parameter estimation in the 20 nHz injections and the lag in constraints returned by the $\mathrm{CW_{ET}}$ model for the 5 nHz injections. In the 20 nHz injections, SP retains a slight edge over $\mathrm{CW_{ET}}$, but both models are largely comparable. The SP model's frequency resolution is limited by the discrete binning, which shifts slightly with increasing array timespans. This creates the sawtooth pattern seen in the SP frequency constraints, wherein the CW is identified, but in each subsequent dataset the frequency bin location shifts slightly.}
    \label{fig:allparamconst}
\end{figure*}

All models lock onto the correct origin region of the sky, as shown in the evolution of the accuracy of parameter recovery in \autoref{fig:allparamconst}, displaying the realization-median 68\% posterior credible regions from each model. However, it is clear that $\mathrm{CW_{FS}}$ achieves the best performance in sky location recovery, stabilizing to the true values early on. As also seen in \autoref{fig:allparamconst}, the binary GW frequency is recovered and constrained by $\mathrm{CW_{FS}}$ first, followed by the other two models--- SP, then $\mathrm{CW_{ET}}$. The $\mathrm{CW_{ET}}$ model struggles particularly in the 5 nHz injections due to the unmodeled interference between Earth and pulsar terms at low frequencies. In the 20 nHz injections, SP and $\mathrm{CW_{ET}}$ both perform similarly in frequency estimation. For all source varieties, $\mathrm{CW_{ET}}$ does recover a tighter constraint than SP by the last time slice due to the latter model's inherent poor frequency resolution, limited as it is due to our linear frequency binning of power. $\mathrm{CW_{ET}}$'s improvement on frequency estimation over SP occurs over a range of source strengths ($6<\mathrm{(S/N)_\Lambda}<19$) depending on the binary parameters.

\subsection{Model Performance}

We now discuss the various expected and unexpected modeling behaviors that we identified in our different analyses. This should be helpful as a benchmark for future studies that explore similar methods or the overarching question of an emerging binary GW signal. 

\begin{figure*}
    \centering
    \includegraphics[width=\textwidth]{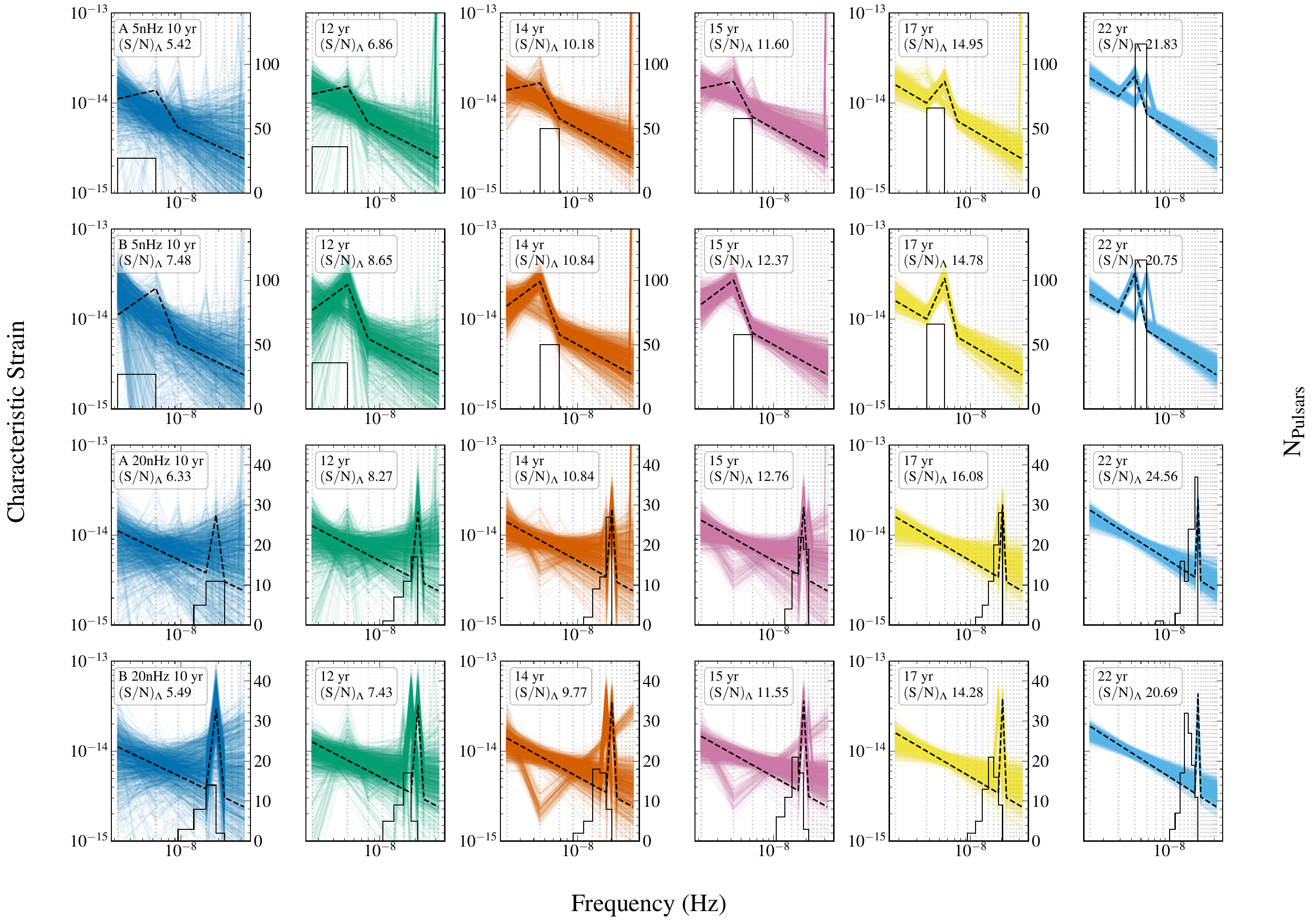}
    \caption{Evolution of the characteristic strain spectrum found by the SP model with an increasing number of pulsars and timing baseline. The injected spectrum is the black dashed line and the black histograms denote the distribution of pulsar term frequencies in the array. Each plot shows one thousand samples from a realization-averaged posterior. In early time slices before the binary is well characterized, the model will sometimes invert the powerlaw used to model the GWB in high frequency binary injections. Posterior support for adjacent bins is indicative of mismodeled spectral leakage and/or pulsar terms.}
    \label{fig:charstrain}
\end{figure*}

Across all models applied to the 20 nHz injections, we find biased estimations toward GWBs with larger amplitudes and shallower spectra, as can be seen in \autoref{fig:allparamconst}. This can be well understood by inspecting the reconstructed characteristic strain spectra of our simulations in the bottom two rows of \autoref{fig:charstrain}. We can see that the true injected spectrum has a high frequency peak due to the CW signal, which can be easily mis-modeled by a shallower power-law with higher amplitude. In fact, this behavior has been seen and characterized previously in other studies of CW searches with background models, namely \citet{2024arXiv240721105F, 2023ApJ...959....9B} and \citet{2024A&A...683A.201V}. We also find that all models recover the frequency more accurately in the 20 nHz case at the final timeslice. This may be due to the fact that the GWB has less power at higher frequencies, translating to less noise for the CW signal to compete with.
We also find that $\rm{CW_{FS}}$ has better binary parameter recovery for the 20 nHz injections. This is discussed in detail in \citet{2025arXiv251001316P}.

We also demonstrate in \autoref{fig:allparamconst} and \autoref{fig:charstrain} that the SP model does indeed gradually constrain the GW frequency to a single frequency bin in which the injected value lies. However, sometimes the posterior distributions for the SP frequency can favor two adjacent frequency bins. This is likely due to covariances between frequencies in our discrete, linear basis applied to a finite timespan of data (i.e., spectral leakage). Yet another potential contribution may come from the distribution of pulsar terms in the array. Together, they can appear as excess power in frequency bins below the binary's Earth-term frequency. Some examples of such double-peaked behavior can be seen in the 12-, 14-, and 17-year panels of the A20 and B20 injection in \autoref{fig:charstrain}. The double peaks in the 22-year panels for A5, B5 in \autoref{fig:charstrain} are likely due to spectral leakage since the source parameters create little difference between the Earth- and pulsar term frequencies (see black histograms of pulsar term frequency distributions).

We find that in late time slices for high frequency injections, the sampler can become trapped in local maxima of the SP likelihood. This region of parameter space corresponds to an inverted power law with the spike placed in the lowest frequency bin. This is a result of model misspecification. The combination of spectral leakage around the frequency bin in which the binary falls, pulsar term effects not captured by the model, and the array's insensitivity at $f=1/\mathrm{year}$, all contribute to the power-law for the GWB model being contorted and ill-fit to produce an inversion of the true, underlying power-law model. When this occurs, the sky localization is usually distributed over large regions of sky or constrained to a single pulsar's antenna pattern. This could be a misleading result, but the inversion of the power law should be indicative of an improper model fit.

In our analyses, the $\mathrm{CW_{ET}}$ model struggled significantly to converge and lock onto the true binary parameters. Even in high $\mathrm{(S/N)_{\Lambda}}$ datasets, the sampler often would only explore local maxima, requiring fine tuning of the sampler or parallel tempering to explore the highest likelihood regions. These were admittedly rich, realistic-format datasets, containing intrinsic pulsar noise processes, a spatially correlated GW background, and the individual binary signal of interest. Missing from these datasets, however, is an extra level of authenticity due to the lack of overlapping signals from a synthesized binary population. This suggests that the $\mathrm{CW_{ET}}$ model would struggle even more in that scenario.

In early tests with the datasets used in this work, we also found that a $\mathrm{CW_{ET}+CURN}$ model is too mis-specified to the injected $\mathrm{CW_{FS}+GWB}$. Recovered parameters were drastically biased due to the $\mathrm{CW_{ET}}$ portion of the model being the only component containing quadrupolar spatial correlations needed to fit the various GW signals present in the data. 
This model could also be impeded by two different pulsar term signal confusion effects, both identified in \citet{2024arXiv240721105F}.
The first confusion effect is the one previously described, wherein low-frequency binaries have pulsar terms that are close in frequency to the Earth term signal being searched for. The interference from these pulsar terms hinders the model's ability to resolve the binary. 
The second pulsar term confusion effect is the attribution of
unmodeled pulsar terms to the $\mathrm{CURN}$ model. This is due to the pulsar terms manifesting as a common, uncorrelated noise process in the array. While the former pulsar term confusion effect would bias CW parameters, the latter would hamper accurate GWB parameter estimation. 

While \citet{2024arXiv240721105F} found that even $\rm{CW_{FS}+CURN}$ models can succumb to the latter pulsar term confusion effect, returning inaccurate GWB parameter estimation,
we do not observe this behavior. 
Evidence of the pulsar terms being assimilated into the CURN model is not present, as it was in \citet{2024arXiv240721105F}. For example, our chirp mass and spectral parameter posteriors are accurate.
Additionally, we find that our $\rm{CW_{FS}+CURN}$ model does recover information from the pulsar terms, as seen in Figure 7 of \citet{2025arXiv251001316P}. 
However, the biases in \citet{2024arXiv240721105F} are found when reweighting CURN posteriors to HD. Since we do not reweight our CURN posteriors (see \autoref{subsec:disc-modimp} for further details), we cannot exclude the possibility that the bias is present. There are also non-negligible differences between our simulated datasets and theirs, namely the PTA configuration and signal strengths. For these reasons we caution against direct comparison between the results of \citet{2024arXiv240721105F} and this work. Both present unique investigations into CW detection in the context of a GWB, and further work to understand CURN biases incurred by pulsar terms is warranted.

\section{Discussion}
\label{sec:Disc}
\subsection{Interpretation of Model Hierarchy}
We first examine why the models show such a consistent hierarchy in parameter estimation and source identification across the suite of injections. We believe this to be the result of a combination of model misspecification and source confusion. $\mathrm{CW_{FS}}$ benefits from utilizing the full signal model, including pulsar terms, when modeling the binary. The only misspecification present is in the CURN approach to modeling the GWB, thereby neglecting HD spatial correlations. 
By contrast, the $\mathrm{CW_{ET}}$ model has the correct model for the GWB, but its signal template lacks pulsar terms. This results in the pulsar term existing as a source of noise that competes with the model. This has adverse effects in the 5 nHz cases, consistent with \citet{2024arXiv240721105F}, where there is minimal frequency evolution between the Earth and pulsar terms. 

Finally, our SP anisotropy technique is the coarsest approach considered in this work. It includes effects of the Earth and pulsar terms stochastically, as well as HD correlations for the GWB. This enables it to outperform $\mathrm{CW_{ET}}$ when the latter is impaired by the pulsar term. For the high frequency binaries, the two models have similar performance, with SP having a slight edge over $\mathrm{CW_{ET}}$. The simplicity and mismatch of the SP model with respect to the injected signal eventually becomes a shortcoming too egregious to ignore, limiting its binary parameter recovery when compared to $\mathrm{CW_{ET}}$. Averaged over injected binary parameters, $\mathrm{CW_{ET}}$ achieves more precise sky localization at $\mathrm{(S/N)_\Lambda} \sim 12$ and improved frequency constraints at $\mathrm{(S/N)_\Lambda} \sim 13$.

Questions have lingered in the literature focusing on whether the simplicity of anisotropy models could retrieve binary parameters earlier than waveform-based searches that have many more parameters \citep{2020PhRvD.102h4039T}, but this work lends credence to the contrary. The signal model that matches as closely to truth ($\mathrm{CW_{FS}}$) performs the best, even with the much larger number of parameters. Many of these parameters are in fact simply filling their priors, such that they do not contribute to any sort of Occam penalty in the Bayesian evidence for the model. 

Nonetheless, the SP model's capabilities will be useful as an avenue for cross-validation in individual binary searches. While the key parameter estimations of sky location and frequency are less constrained than the full signal model, consistent parameter estimations would lend confidence to results from a $\mathrm{CW_{FS}}$ search. Beyond cross-validation, SP is the most well-equipped anisotropy technique for finding the narrowband, small angular scale anisotropy predicted by SMBHBs.
SP is faster to converge than other anisotropy models due to its narrowband frequency dependence and its simplicity.
SP adds only four additional parameters compared to an isotropic GWB search. The widely-used spherical harmonic and square-root spherical harmonic models require $(l_{max}+1)^2-1$ parameters per frequency bin probed to characterize anisotropy with angular scales $\theta = 180^{\circ}  / l_\mathrm{max}$ \cite{2020PhRvD.102h4039T, 2023ApJ...956L...3A}. Thus, a great number of parameters are required to explore small-scale anisotropies associated with individual SMBHBs. This tradeoff between the number of parameters and the angular scale being probed does not exist with SP, which is a key benefit of the model. An additional hindrance facing other anisotropy techniques is that we do not know which frequency bin the binary falls in a priori. Therefore, we would need to use an alternative approach to pick which frequency bin to search for anisotropy.
This could require either a preliminary analysis to find where there are power-law departures, or a full frequency-dependent anisotropy analysis. The latter search would require $\mathrm{N_{frequencies}}$ runs to fully describe the data.

\subsection{Caveats of the Simulations}
While our results may seem to imply that the simpler SP and  $\mathrm{CW_{ET}}$ models offer no real usefulness in CW searches, it is critical to remember the style of injections used in this work. The simulated datasets are still idealistic, i.e., a Gaussian power-law background with one resolvable SMBHB. A realistic population of SMBHBs would introduce more signal complexity and difficulties for each model. Such a background would have more spectral fluctuations, especially at higher frequencies, due to the potential for multiple individually resolvable binaries. With multiple peaks in the background, the SP model would encounter a multimodal likelihood surface that could inhibit its detection capabilities and parameter estimation. Given enough data, the strongest binary may be recovered, depending on its relative strength to the other individual sources and their respective frequencies. However, multiple threshold binaries in a single frequency bin could confuse the model. Problems erupting from overlapping signals are not unique to SP. Indeed, the incoherent superposition of responses from different binaries can confuse a full signal template model. \citet{2025arXiv250216016G} found that this model can fail to recover the loudest binary or return inaccurate parameter estimation in some scenarios. 

A solution to these issues is through transdimensional signal modeling using Reversible Jump MCMC (RJMCMC), which empowers the data to inform the number of favored resolved sources. Such an approach has been developed for a full signal template \citep{2020CQGra..37m5011B}; however, a transdimensional anisotropy search, tuned for CW detection, has not. The closest to a SMBHB-population targeted anisotropy model is the multi-pixel analysis performed in \citet{2020PhRvD.102h4039T}. However, that work used product-space sampling with increasingly disjoint models as more binaries were being modeled, rather than an RJMCMC algorithm. The development of transdimensional GWB anisotropy approaches is important and will be the focus of future work. 
Whether the model-efficacy hierarchy persists in a realistic background case remains to be seen and will also be investigated in the future.

Lastly, the simulated datasets contained a GW signal from a circular binary; however, the current GWB spectral measurements could suggest non-negligible eccentricity \cite{2024MNRAS.532..295F, 2022MNRAS.511.4753G, 2024A&A...691A.212R}. \citet{Taylor2016} explored the penalties incurred by searching for an eccentric binary with an Earth term template model, finding that it performs reasonably well for binaries with low and very high eccentricities. How well a circular, full-signal model would perform is an open question. SP could struggle in an eccentric binary scenario as the GW emission would be distributed into higher harmonics, reducing the power in a single frequency bin. \citet{2025arXiv250614882E} and \citet{2025ApJ...993..118S} exploit this effect, using the covariance between anisotropy in different frequency bins to study the eccentricity distribution of the GWB. At very high eccentricity, when the GW timing-residual power peaks at the orbital frequency, the SP model would be less affected. This is because GW power would be concentrated in one frequency bin, corresponding to the orbital frequency, rather than distributed broadly in the case of middling eccentricity binaries. These considerations likely apply only to low-frequency binaries, as GW emission will circularize the binary \citep{1963PhRv..131..435P}. 

\subsection{Model Improvements}
\label{subsec:disc-modimp}
There are improvements that can be made to the SP model. One shortcoming of our approach is its coarse frequency basis. Its current implementation shares the frequency basis used for modeling the GWB, leaving it with frequency resolution equal to $\Delta f =1/T$. This leaves the model susceptible frequency bin covariances (i.e. spectral leakage) due to the finite observation window and the inherent discrete and linear nature of the basis (See 22-year panels of A5 and B5 injections in \autoref{fig:charstrain}). Moreover, many pulsar term frequencies could fall in the same frequency bin, producing excess power that could draw SP away from the binary's frequency bin (See 12-, 14-, and 17-year panels of A20 and B20 injections in \autoref{fig:charstrain}). This could be improved by explicitly accounting for data windowing through a sinc function model rather than a single-bin spike in the power spectrum \citep{2025arXiv250613866C}. This would enable the model to have higher frequency resolution and mitigate spectral leakage confusion. 

The SP model also assumes equal power from the plus and cross polarizations of GWs, effectively producing an unpolarized process. This contributes to the misspecification of the model with respect to an elliptically polarized CW signal. It is possible to develop a polarized version of SP, containing a $\rho_{\rm SP+}$ and a $\rho_{\rm SP\times}$ parameter with common sky location and frequency. In idealized tests, this model recovered a difference between the two polarization amplitudes only at high S/N ($\mathrm{(S/N)_{opt} \approx 20-50}$), where template models would easily outperform SP anyway.

Despite the leading performance of the $\mathrm{CW_{FS}}$ model, our analyses are conducted with \texttt{QuickCW}, which does not incorporate the full details of the injected signals. That is, \texttt{QuickCW} employs the simpler CURN model for the GWB, excluding HD spatial correlations. Biased parameter recovery due to model misspecification is inevitable in this case, as seen in \autoref{fig:allparamconst}. The GWB parameters are mildly biased, while the CW parameters are generally not. To mitigate these biases, it will be necessary in future work to include HD correlations in the $\mathrm{CW_{FS}}$ model \citep{2023ApJ...959....9B,2024arXiv240721105F,2024A&A...683A.201V}.

Such analyses are computationally prohibitive, though, especially as the number of pulsars and observations increase. As a workaround, one can use post-processing methods, such as sample reweighting, to obtain posteriors corresponding to an HD-correlated likelihood \citep{2023PhRvD.107h4045H}. However, we find that reweighting CURN+CW$_{\mathrm{FS}}$ posteriors to HD+CW$_{\mathrm{FS}}$ posteriors results in very low efficiency, or high resampling error. As explained in \citet{2023PhRvD.107h4045H}, the error increases when the posterior distribution of the approximate model (CURN+CW$_{\mathrm{FS}}$) has minimal overlap with the posterior distribution of the target model (HD+CW$_{\mathrm{FS}}$). The lack of overlap ultimately leads to a poorly-reconstructed target posterior. Because of this, we do not account for HD correlations via reweighting.

The low reweighting efficiency could arise for a few reasons. Our simulated datasets quickly grow in complexity from one time slice to the next: the increase in the number of pulsars ($\mathrm{N_{psr}}$) begets an increase of 2$\mathrm{N_{psr}}$ RN parameters (power-law amplitude and spectral index per pulsar). \citet{2023PhRvD.107h4045H} note that when comparing CURN and HD posteriors through the Kullback–Leibler (KL) divergence \cite{kullback1951information}, a larger divergence results in lower sampling efficiency. In particular, pulsar RN parameters are responsible for the divergence more so than the GWB parameters. Similar to RN parameters, 2$\mathrm{N_{psr}}$ CW parameters are added (pulsar distance and pulsar term phase), making the likelihood surface increasingly demanding to explore. The CW signals we inject are also fairly strong and become more prominent at later time slices. Unsurprisingly, we find that the reweighting efficiency decreases with each subsequent slice. For example, the 10-year slice, $\mathrm{(S/N)_\Lambda}$~$>$~5, has an average efficiency of $\lesssim 1\%$ and quickly falls to $\ll 1\%$ beyond this slice. It is therefore likely that the CURN+CW$_{\mathrm{FS}}$ and HD+CW$_{\mathrm{FS}}$ posteriors are not similar enough for the reweighting technique to be effective here \citep{2023ApJ...951L..50A}. The technique will also become increasingly limited as real PTA datasets become more sensitive in the future.

New strategies for modeling the full signal structure of both the GWB and the CW are crucially needed. For instance, \citet{2024arXiv241213379G} present a step in this direction, taking a Fourier basis approach to deterministic signals that permits rapid CW searches alongside HD-correlated GWB modeling. When computationally feasible, it would be interesting to compare the efficacy of a $\mathrm{CW_{FS}}$ model which includes HD correlations against the SP and $\mathrm{CW_{ET}}$ models. The $\mathrm{CW_{FS}}$ and HD GWB models have previously been seen to compete for signal power \citep{2023ApJ...951L..50A,EPTA_InPTA_DR2_cw}, putting into question whether the $\mathrm{CW_{FS}}$ model would still detect and characterize the CW signal as early as it does in this work.

\section{Conclusions}
\label{sec:conclusion}
This work was motivated by an unsettled question from \citet{2020PhRvD.102h4039T}: will a single SMBHB reveal itself initially through anisotropy? We have attempted to address this question by creating a physically-motivated model for frequency-resolved anisotropy, searching for an individual SMBHB via its cross correlations, and comparing it to existing template-based searches. 

We created $40$ simulated PTA datasets that were inspired by the scale and structure of the forthcoming IPTA Data Release 3. This was not with the goal of predicting what may be found in the latter, but rather to be representative of the noise and timing heterogeneity of a realistic, near-future PTA dataset. Each of the $40$ datasets were broken into 11 cumulative time segments, emulating the growth of PTA datasets in time and pulsars, and containing a Gaussian GWB along with a single resolvable SMBHB. To every dataset, we apply two contemporary template-based CW search pipelines, as well as our novel anisotropy analysis, to compare their detection and parameter recovery abilities. 

We find that a full CW signal template model, $\mathrm{CW_{FS}}$, containing both Earth and pulsar term contributions to the binary signal, outperforms the \textit{Spike Pixel} (SP) anisotropy search as well as an Earth-term-only template search ($\mathrm{CW_{ET}}$). In terms of detection statistics, the $\mathrm{CW_{FS}}$ model attains higher Bayes factors for equivalent time slices than other methods, despite the significantly greater dimensionality of its parameter space. Moreover, $\mathrm{CW_{FS}}$ shows superior parameter recovery over the other two methods. It constrains key parameters for EM follow-up (sky location, GW frequency) more precisely and at lower signal-to-noise ratio than the other approaches examined here. 

Inevitably, the Savage-Dickey technique with which we computed Bayes factors became unreliable. This is expected as the MCMC sampling coverage of the CW amplitude's posterior tail should become poorer due to increasing signal strength. We then used model-specific likelihood ratios as a proxy detection statistic to assess the performance of our various models in the higher signal-to-noise regimes. 

These ratios qualitatively agree with our Bayesian results, showing the greater ability of the deterministic models over a model for anisotropy. Initially, however, the SP anisotropy method recovers higher Bayes factors and improved parameter estimation over the $\mathrm{CW_{ET}}$ in low-frequency injections. This is due to the minimal frequency evolution between Earth and pulsar terms, seen also by \citet{2024arXiv240721105F}, whose interference causes confusion that inhibits the $\mathrm{CW_{ET}}$ model's performance. In the case of binaries with a GW frequency of 20 nHz, SP and $\mathrm{CW_{ET}}$ are comparable in Bayes factors and parameter recovery at early time slices. Since the $\mathrm{CW_{ET}}$ model is determinsitic and closer to the true GW signal model, it ultimately surpasses the simplistic SP model, returning more precise and accurate parameter estimation (see \autoref{subsec:res-paramest}). The delayed performance of $\mathrm{CW_{ET}}$ can be explained by the combination of several factors, originating from intrinsic pulsar spin processes, unmodeled pulsar terms, and the underlying GWB, which cloud the model's abilities until the higher signal-to-noise regime. MCMC exploration was also very challenging for the $\mathrm{CW_{ET}}$ model, and further use will require, e.g., fine-tuning of proposal distributions, and parallel tempering for reliable convergence. 

Thus, our study here has shown that the full CW signal model achieves the best detection and characterization metrics for an emerging single SMBHB in PTA data. Yet, despite the realistic format and structure of our PTA configuration, the signal injections were still rather idealized; they contained a Gaussian GWB with a power-law strain spectrum, with a single CW signal on top. Further study is needed in which a population of discrete SMBHBs produces overlapping GW signals in PTA datasets. There are still no data-analysis strategies for PTAs that can rapidly search for multiple CW signals in tune with an HD-correlated GWB and PTA noise in a transdimensional model selection scheme. Hence, the SP model's usefulness may be twofold: as a cross-validation technique of other more refined approaches, and for rapid transdimensional identification of threshold SMBHB signals emerging at multiple GW frequencies. 
We plan to investigate this in future work by developing a transdimensional \textit{Spike Pixel} anisotropy model, in a bid to search for continuous gravitational waves emerging from a realistic GW background composed of many overlapping SMBHB signals.

\begin{acknowledgments}
LS and PP would like to thank Kyle Gersbach, William Lamb, Celia Fielding, and Matt Miles for their ideas and helpful discussions. LS would also like to thank Jaelyn Roth for discussions surrounding final parsec solutions. LS is forever indebted to Dr. Juliet Liss and Dr. Alexander Schult for raising them and impressing upon them the importance of education. Without these thoughtful efforts, this work would not exist. We would also like to thank Bence Bécsy and Caitlin Witt for productive conversations around the efficiency of reweighting CW+GWB models. We would like to further thank Bence Bécsy and Irene Ferranti for thoughtful comments on a draft of this manuscript. PP acknowledges support from NASA FINESST grant number 80NSSC23K1442. LS is greatly appreciative of support from NSF AST2307719 and NRT-2125764. SRT acknowledges support from an NSF CAREER \#2146016, NSF AST-2307719, NSF NRT-2125764, and NASA LPS-80NSSC26K0342. SRT also acknowledges support from a Chancellor's Faculty Fellowship from Vanderbilt University. M. C. acknowledges support by the European Union (ERC, MMMonsters, 101117624). The authors are members of the NANOGrav collaboration,
which receives support from NSF Physics Frontiers Center award number 1430284 and 2020265. This work was conducted in part using the resources of the Advanced Computing Center for Research and Education (ACCRE) at Vanderbilt University, Nashville, TN. This work utilized the software suites \texttt{astropy} \cite{astropy:2013, astropy:2018, astropy:2022}, \texttt{corner} \cite{corner}, \texttt{HEALPix} \cite{2005ApJ...622..759G}, \texttt{healpy} \cite{Zonca2019}, \texttt{Jupyter} \cite{soton403913}, \texttt{La Forge} \cite{2020zndo...4152550S}, \texttt{Matplotlib} \cite{Hunter:2007}, \texttt{Numpy} \cite{harris2020array},  and \texttt{Scipy} \cite{2020SciPy-NMeth}.
\end{acknowledgments}


\bibliography{apssamp}

\end{document}